\documentclass{aa}  

\usepackage{graphicx}
\usepackage{subcaption}
\usepackage{booktabs}
\usepackage{amsmath}
\usepackage{lipsum}
\usepackage{comment}
\usepackage{adjustbox}
\usepackage{multirow}
\usepackage{txfonts}
\usepackage{float}
\usepackage{natbib}
\usepackage[hypertexnames=false]{hyperref}
\usepackage{lscape}
\newcommand{\civ}{\texorpdfstring{\ion{C}{iv}}{C IV}}
\newcommand{\nv}{\texorpdfstring{\ion{N}{v}}{N V}}
\newcommand{\mgii}{\texorpdfstring{\ion{Mg}{ii}}{Mg II}}
\newcommand{\siiv}{\textup{Si\,\textsc{iv}}}
\newcommand{\cii}{\textup{[C\,\textsc{ii}]}}
\newcommand{\ciiALLOWED}{\textup{C\,\textsc{ii}}}

\newcommand{\alii}{\textup{Al\,\textsc{ii}}}
\newcommand{\aliii}{\textup{Al\,\textsc{iii}}}
\newcommand{\feii}{\textup{Fe\,\textsc{ii}}}
\newcommand{\siii}{\textup{Si\,\textsc{ii}}}
\newcommand{\oii}{\textup{O\,\textsc{ii}}}
\newcommand{\mgi}{\textup{Mg\,\textsc{i}}}
\newcommand{\lya}{\textup{Ly$\alpha$}}
\newcommand{\hb}{\textup{H$\beta$}}

\begin{document}

   \title{Multi-phase investigation of outflows in the circumgalactic and interstellar media of luminous quasars at z$\sim$5}

   \author{M. Brazzini\inst{1,2,3}, V. D'Odorico\inst{2,3,4}, M. Bischetti\inst{1,2}, C. Feruglio\inst{2,3}, G. Cupani\inst{2,3}, G. Becker\inst{5}, R. Tripodi\inst{3,6} 
          }

   \institute{
   Department of Physics, Astronomy Section, University of Trieste, Via G.B. Tiepolo, 11, I-34143 Trieste, Italy \email{matilde.brazzini@phd.units.it} 
   \and INAF - Osservatorio Astronomico di Trieste, Via G. B. Tiepolo 11, I-34143 Trieste, Italy \email{matilde.brazzini@inaf.it} 
   \and IFPU - Institute for Fundamental Physics of the Universe, Via Beirut 2, I-34014 Trieste, Italy 
   \and Scuola Normale Superiore, P.zza dei Cavalieri, I-56126 Pisa, Italy
   \and Department of Physics \& Astronomy, University of California, Riverside, CA 92512, USA
   \and Faculty of Mathematics and Physics, University of Ljubljana, 19 Jadranska ulica, Ljubljana, 1000, Slovenia}

  \abstract
   {}
   {Outflows from active galactic nuclei are invoked as the principal feedback process regulating the co-evolution of supermassive black holes and their host galaxies. 
   Because of their multi-phase and multi-scale nature, an exhaustive description of these winds should exploit multiple tracers. However, connecting various outflow features remains a challenge. 
   The aim of this work is to provide a complete characterisation of outflows in a sample of $z \sim 5$ quasars, by exploiting the combination of different emission and absorption tracers. 
   }
   {We analysed the UV/optical and FIR continuum, line emission, and absorption in a sample of 39 $z \sim 5$ quasars observed with VLT/X-Shooter and ALMA (available for six objects). 
   We identified broad and narrow absorption lines associated with the quasar and emission lines to determine black hole masses and bolometric luminosities. 
    }
   {Our sample encompasses massive ($M_{\text{BH},\text{\mgii}} = 10^{8.5-10} M_\odot$) and luminous ($L_{\text{bol}} = 10^{46.9-48}$ erg/s) quasars at redshift $5 - 5.7$. 
   They display powerful ionised outflows detected in both emission and absorption, with velocities exceeding 48,000 km s$^{-1}$ in some cases, and lie above the local black hole -- host galaxy mass relation, exhibiting a behaviour similar to that of $z \gtrsim 6$ quasars.  
   These findings suggest a phase of efficient black hole feedback occurring at redshift $z \gtrsim 6$ and likely persisting down to $z \sim 5$, characterised by rapid black hole growth exceeding that of the host galaxy.
   The fraction of quasars with outflow detections in absorption is higher for larger \civ-\mgii\ velocity shifts, suggesting that while the physical mechanisms powering the two outflow phenomena detected in emission and absorption may differ, a correlation exists between them.
   }
   {}


    \titlerunning{Multi-phase investigation of outflows in the ISM and CGM of luminous quasars at z$\sim$5 }
    \authorrunning{M. Brazzini et al.}
    
   \maketitle

\section{Introduction}
The co-evolution of supermassive black holes (SMBHs) and their host galaxies is regulated by complex feedback mechanisms involving multi-phase and multi-scale outflows, which originate near the central SMBH and extend up to kiloparsec scales \citep{fabian_observational_2012, harrison_observational_2024}. 
Especially in luminous quasars, such outflows are thought to make a major contribution by significantly impacting the host galaxy interstellar and circumgalactic media (respectively ISM and CGM; e.g. \citealp{costa_environment_2014}). 
They inject into the ISM both energy and momentum, and are responsible for heating and expelling gas towards the CGM and even the intergalactic medium (IGM), further regulating gas accretion onto the galaxy with crucial consequences on both SMBH growth and
star formation. 

Outflows can be identified via different indicators, spanning from asymmetric blue-shifted wings in emission lines to blue-shifted absorption lines against quasar continuum emission.
Both emission and absorption features can occur with various degrees of ionisation and in different spectral bands. 
The colder molecular and neutral phases are typically observed in the rest-frame far-infrared/sub-millimeter, while the ionised phase is primarily traced in the optical/UV.
Outflow detections in emission have been reported, for example, in \citealp{feruglio_quasar_2010} for the molecular phase, \citealp{fan_alma_2018} for the neutral phase, and \citealp{carniani_ionised_2015} and \citealp{bischetti_wissh_2017} for the ionised phase. 
Outflow detections in absorption can be found among many works: \citealp{fischer_herschel-pacs_2010, butler_molecular_2023} for the molecular phase, \citealp{cicone_very_2015} for the neutral phase, and \citealp{culliton_probing_2019} and \citealp{bischetti_fraction_2023, bischetti_multiphase_2024} for the ionised phase.   
At low and intermediate redshifts ($z<2-3$), there are few studies \citep{feruglio_multi-phase_2015, perna_x-raysdss_2017, fiore_agn_2017, herrera-camus_molecular_2019, herrera-camus_agn_2020,xu_evidence_2020, fluetsch_properties_2021, riffel_agnifs_2023, speranza_multiphase_2024} reporting the detection of all the different ionised, neutral and molecular outflow phases within the same source. It emerges that such multi-phase winds are a ubiquitous phenomenon in active galactic nuclei (AGNe), with the molecular phase being slower and more compact than the ionised one, and carrying the bulk of the outflow mass.  
At higher redshifts, the multi-phase characterisation of outflows becomes more difficult, although strong AGN-driven winds are expected in response to the large radiative outputs of highly accreting SMBHs found at $z\sim 6$ \citep{ zappacosta_hyperluminous_2023, bischetti_multiphase_2024}. 
In particular, the high-velocity line wings that trace outflows in emission are difficult to detect, especially for the cooler and slower phases.
A possible solution is stacking multiple spectra to improve their sensitivity \citep{bischetti_widespread_2019}.  
Alternatively, one can resort to absorption as the principal investigation tool
for studying outflows at $z \gtrsim 2-3$.

Absorption lines are usually classified based on the full width at half maximum (FWHM) of their profiles. 
We distinguish between broad absorption lines (BALs), characterised by deep, broad absorption troughs with FWHM $> 1000-2000$ km s$^{-1}$, and narrow absorption lines (NALs), with FWHM $\lesssim 300$ km s$^{-1}$. 
Systems with intermediate FWHM values are commonly denoted as mini-BALs. 

While BALs are unambiguously associated with black hole-driven winds, their broad and often saturated profiles make it challenging to derive the physical conditions of the gas.
In contrast, although their physical origin is more uncertain, as they can arise from a variety of gaseous environments, the significance of NALs lies in the fact that prominent UV doublets, such as \civ\,$\lambda \lambda 1548,1550$ \AA, \nv\,$\lambda \lambda 1238, 1242$ \AA, \siiv\,$\lambda \lambda 1393, 1402$ \AA, and \mgii\,$\lambda \lambda 2796, 2803$ \AA, are generally not blended and can be analysed to determine column densities, ionisation levels, and metallicities of IGM, CGM, and even host galaxy ISM (e.g. \citealp{rudie_column_2019, nunez_kbss-inclose_2024}). 
If located at small velocity separations ($\lesssim 10,000$ km s$^{-1}$, in the so-called proximity zone) from the quasar systemic redshift $z_{em}$, NALs likely trace the host galaxy CGM and IGM, and the presence of AGN-driven outflows \citep{ganguly_origin_2001, dodorico_chemical_2004, hamann_sabra_2004, misawa_census_2007, nestor2008, culliton_probing_2019, stone_narrow_2019, perrotta_nature_2016, perrotta_hunting_2018}.
In these cases, we call these lines `intrinsic', as they originate from systems that are physically related to the central AGN. 
However, NALs may also originate from intervening systems intersecting the line of sight. 
Such intervening absorbers are ubiquitous in quasar spectra and are not restricted to a specific redshift interval. 
When considering absorption systems in the quasar proximity zone (collectively denoted as associated systems), the major difficulty lies in discerning the intrinsic absorbers from the intervening absorbers. 
To date, the intrinsic nature of associated absorption lines is inferred via some general criteria based, for example, on their time variability, their broad profiles (although with FWHM never exceeding a few hundred kilometers per second), the presence of excited-state fine-structure lines, partial coverage effects, and/or the presence of high ionisation transitions that are not found in intervening absorbers. 
Among the highly ionised absorption species commonly used to identify ionised outflows, such as \civ, \siiv\ and \nv, \cite{perrotta_nature_2016} found that \nv\ is the ion that best traces the effects of the quasar ionisation field, thus offering the best statistical tool to identify intrinsic systems. 
In a subsequent work, \cite{perrotta_hunting_2018} confirmed this result by detecting a statistically significant \nv\ absorption signal only within $\sim 5000$ km s$^{-1}$ from $z_{em}$. 

In this work, we analyse a sample of 39 quasars at $z = 4.99-5.70$ observed with the echelle spectrograph X-Shooter on the Very Large Telescope (VLT). 
We systematically search for \civ\ and \nv\ NAL systems lying in the quasar proximity zone, to detect and characterize ionised outflows in absorption. 
Our analysis is complemented by a study of BALs and an investigation of the quasar UV emission spectrum and far-infrared (FIR) properties, particularly the \cii158$\mu$m emission, using available archival data from the Atacama Large Millimeter/submillimeter array (ALMA).
The primary goal of this combined approach is to link the properties of ionised outflows detected in absorption, through either NALs or BALs, with other galaxy and black hole properties inferred from both UV and FIR emission lines.  
In fact, from the UV \civ\ and \mgii\ emission lines, we can constrain the central black hole mass, quasar bolometric luminosity, and Eddington ratio, while from the \cii\ 1D emission profile and 2D emission map, we can estimate the \cii\ luminosity, its physical extension, and the host galaxy dynamical mass. 

In the local Universe, the co-evolution of SMBHs and their host galaxies manifests itself in the existence of scaling relations between SMBH masses and various galaxy properties, for example the host galaxy bulge mass $M_{\text{host}}$, its luminosity $L_{host}$, and stellar velocity dispersion $\sigma_\star$ \citep{ferrarese_fundamental_2000, tremaine_slope_2002, marconi_relation_2003, mcconnell_revisiting_2013, kormendy_coevolution_2013}. 
The validity of these relations at high redshift is still under debate; interestingly, there is some evidence of their variation with cosmic time, especially for the $M_{\text{BH}} -M_{\text{host}}$ relation \citep{peng_probing_2006, decarli_quasar_2010, wang_probing_2016, bischetti_wissh_2018, pensabene_alma_2020, tripodi_hyperion_2024}. 
In particular, high-redshift ($ z \gtrsim 2-3$) SMBHs appear to lie above the local $M_{\text{BH}}-M_{\text{host}}$ relation, suggesting that at early epochs, black hole growth must have preceded that of the host galaxy. 
The potential evolution in the interaction between SMBHs and host galaxies would then be a consequence of the evolution in the properties of outflow and their impact on these galaxies.
Within this context, it is crucial to exploit all available outflow tracers to gather comprehensive information about these powerful yet poorly understood phenomena, especially at $z \gtrsim 2$. 
Our statistical analysis of associated and intrinsic NALs aims to fill this gap and provide insights into these absorption features that are ubiquitous in quasar spectra but remain underused due to their incomplete understanding. 

This paper is organised as follows. 
In Section \ref{Section-X-Shooter-data-analysis} we present the X-Shooter dataset. We then describe the data reduction (Section \ref{section-x-shooter-data-reduction}), the identification, and fit of NALs (Section \ref{section-x-shooter-absorption-data}) and their classification (Section \ref{section-x-shooter-construction-absorber-sample}). 
We present the modelling of UV continuum and emission lines in Section \ref{section-x-shooter-modeling-uv-spectra} and the BAL identification procedure in Section \ref{section-x-shooter-BALs}. 
In Section \ref{section-ALMA-data} we introduce the ALMA data, and describe their reduction procedure. 
We present our absorption analysis results in Section \ref{section-results-statistics}, while the emission line  results are presented in Section \ref{section-results-quasar-properties-emission-lines} for both the UV (Sect. \ref{section-results-uv_analysis}) and the FIR \cii\ lines (sect. \ref{section-results-cii-emission-and-properties}). 
In Section \ref{section-discussion}, we compare our sample of \civ+\nv\ absorption lines with previous literature results (Sect. \ref{section-discussion-comparison-with-previous-literature-absorption-catalogues}) and investigate the relation between host galaxy and black hole masses (Sect. \ref{section-discussion-coevolution}). 
Lastly, in Section \ref{section-discussion-combining-absorption-emission}, we integrate our absorption and emission analyses. 
Throughout the paper, we assume a $\Lambda$CDM cosmology with $H_0 = 67.3$ km s$^{-1}$ Mpc$^{-1}$, $\Omega _\Lambda = 0.68$, and $\Omega _M = 0.32$ \citep{planck_collaboration_planck_2020}. 

\begin{figure*}
    \centering
    \includegraphics[width=0.9\linewidth]{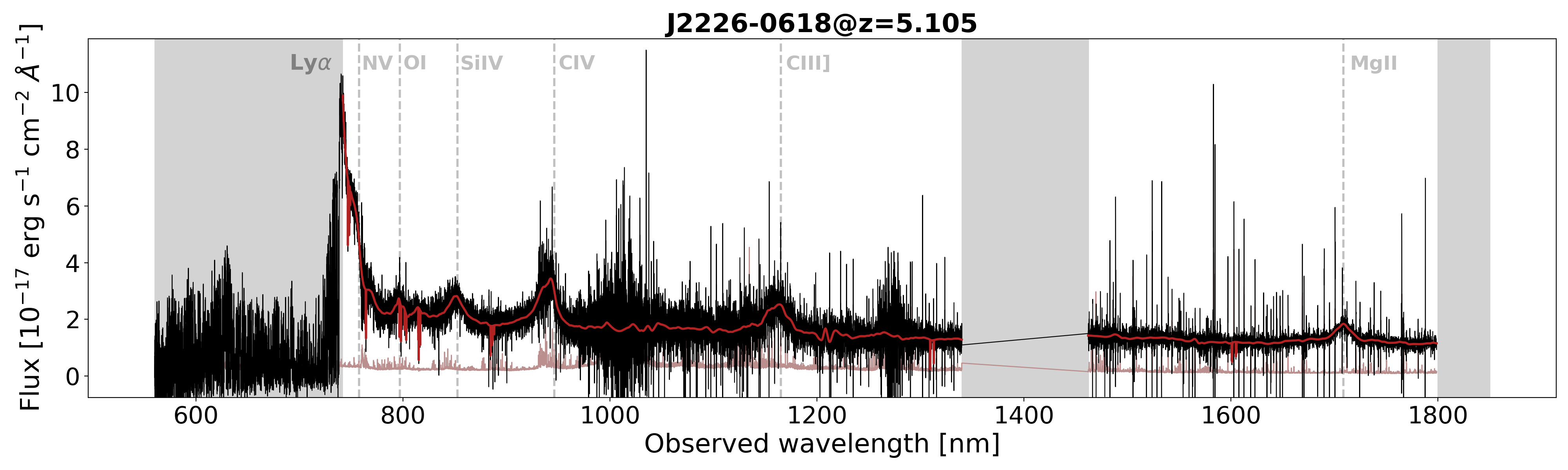}
    \caption{Example of a typical X-Shooter spectrum from our quasar sample, covering the 560--1850 nm observed range, reported in black. In pink we display the associated error. The red curve comprises the \textsc{Astrocook} continuum and emission lines best fit, as well as the identified absorption systems. 
    We label the main emission lines typically identified in quasar spectra.
    The \lya\ forest and telluric-contaminated spectral regions are masked by gray-shaded regions and not fitted. }
    \label{fig:x-shooter-spectrum}
\end{figure*}

\section{X-Shooter data analysis}
\label{Section-X-Shooter-data-analysis}

We consider 39 quasars at $4.99 \lesssim z \lesssim 5.70$ observed using the X-Shooter spectrograph \citep{vernet_x-shooter_2011} at the Very Large Telescope (VLT).
These objects are retrieved from the programmes 098.A--0111 and 0100.A--0243 (PI: Rafelski, M.), which were originally designed to conduct a blind search for DLAs at $z \gtrsim 5$, without any prior information on intervening absorption systems. 
This ensures that these data constitute a large and likely unbiased sample of NALs. 
Despite the attempt to avoid BAL quasars, some survived the original selection procedure and are still found in the sample (Section \ref{section-x-shooter-BALs}). 
For the instrument setup, the 0.9x11" and 0.6x11" slits are adopted for the VIS (550--1020 nm) and NIR (1000--2480 nm) arms, respectively, providing a nominal resolving power of 8900 and 8100. 

\subsection{Data reduction}
\label{section-x-shooter-data-reduction}

The X-Shooter data reduction process is described in detail in \cite{becker_evolution_2019}; here, we provide a brief summary of the key steps and outcomes. 
For each two-dimensional spectral exposure, sky and dark subtraction were performed, followed by the application of telluric absorption corrections.
Then the one-dimensional spectrum was extracted, and a relative flux calibration was performed using a response function generated from a standard star. The absolute flux calibration is carried out at a later stage, and is described in detail in Section \ref{section-results-uv_analysis}.
The final outcome of the reduction pipeline is a one-dimensional spectrum for each of the two considered X-Shooter spectroscopic arms, VIS and NIR, where all exposures are simultaneously combined to optimise the rejection of bad pixels.
The two spectra are re-binned in bins with fixed width of 10 km s$^{-1}$, a fine sampling suitable for the research of narrow absorption features.
The signal-to-noise ratio (S/N) per pixel of 10 km s$^{-1}$ ranges from 4 to 34, with a median of 13, at rest-frame wavelength of 1285 \AA.

\subsection{Identification and fit of metal narrow absorption lines}
\label{section-x-shooter-absorption-data}

\begin{table}[]
    \centering
    \caption{Absorption lines studied in this work. Lines from the same ion are arranged from the strongest to the faintest, corresponding to the highest to the lowest oscillator strengths.}
    \adjustbox{max width=0.45\textwidth}{
    \begin{tabular}{cc}
    \toprule
       Ion  & Transitions \\
    \midrule
       \civ  & $\lambda 1548.204$ \AA, $\lambda  1550.781$ \AA \\
       \nv & $\lambda 1238.821$ \AA, $\lambda  1242.804$ \AA \\
       \siiv & $\lambda 1393.760$ \AA, $\lambda  1402.773$ \AA \\
       \mgii & $\lambda 2796.354$ \AA, $\lambda  280.531$ \AA \\
       \oii & $\lambda 1302.168$ \AA \\
       \siii & $\lambda 1260.422$ \AA, $\lambda  1304.370$ \AA, $\lambda  1526.707$ \AA \\
       \feii & $\lambda 2382.765$ \AA, $\lambda  2600.173$ \AA, $\lambda  2344.214$ \AA, $\lambda  2586.650$ \AA \\
       & $\lambda  1608.451$ \AA, $\lambda  2374$ \AA, $\lambda  1260.533$ \AA, $\lambda  1611.200$ \AA \\
       \alii &  $\lambda 1670.789$ \AA \\
       \mgi & $\lambda 2852.963$ \AA \\
       \aliii &  $\lambda 1854.718$ \AA, $\lambda 1862.791$ \AA \\
    \bottomrule
    \end{tabular}}
    \label{tab:absorption_lines}
\end{table}

From a practical point of view, all the operations of spectra manipulation, continuum fitting and line identification described in this Section are performed using \textsc{Astrocook}\footnote{\url{https://das-oats.github.io/astrocook/}} \citep{cupani_astrocook_2020, cupani_astrocook_2022}, a versatile and interactive Python software environment for the analysis of quasar spectra. 

The first step of our data analysis procedure consists in combining VIS and NIR one-dimensional spectra into a single spectrum. 
To achieve this, we first scale the NIR arm to match the VIS arm over their overlapping range, 1000--1020 nm. 
Next, we cut the two spectra at at 1013 nm, and stitch them together. 
This wavelength is chosen as it marks the point where the errors in the red end of the VIS arm and the blue end of the NIR arm are equal.  
We keep the 10 km s$^{-1}$ original binning on the two spectra for the NAL analysis. 
We set the final wavelength coverage to 560--1850 nm, excluding spectral regions bluewards of 560 nm, to avoid the noisy blue end of the VIS arm, and redwards of 1850 nm, where no relevant spectral features are present. 
An example of a typical spectrum from our sample is reported in Fig. \ref{fig:x-shooter-spectrum}. 

The successive step is the continuum fit, performed with \textsc{Astrocook} using the \textsc{flux\_clip} recipe, which basically: 1) calculates the running median flux in windows of fixed velocity width, 80 km s$^{-1}$ in our case; 2) iteratively identifies and excludes negative outliers with depth $> \kappa$ times the flux error at that wavelength, with the depth estimated with respect to the running median and $\kappa$ a constant to be fixed; 
3) applies a gaussian smoothing to the spectrum to suppress stochastic fluctuations associated with Poisson noise; 4) finally fits a spline to the spectral data. 
We impose $\kappa = 5$ and fix at 200 km s$^{-1}$ the standard deviation for the gaussian smoothing. 
This procedure gives good initial results, which we manually refine through visual inspection using the \textsc{Astrocook} graphical user interface (GUI). 
Additional refinement is particularly needed in regions where the spectrum has a significant slope or is highly variable because of strong telluric and sky lines residuals, as well as around strong and/or clustered absorption lines which tend to lower the continuum when they occupy almost the whole window for the running median estimate. 
Telluric contamination is particularly significant at observed wavelengths of 1350--1450 nm and 1800--1950 nm, in the NIR part of the spectrum: to address this, we mask these regions and exclude them from further absorption analysis.
Lastly, for those objects that present BAL features, we adjust the spectral continuum to follow the BAL trough, as at this point of the analysis our intent is to detect NALs and not to model the BAL troughs, a task we successively addressed (see Section \ref{section-x-shooter-BALs}). 

The procedure described so far works well for spectral regions redward of the \lya\ emission line. 
However, it fails in fitting the quasar continuum at lower wavelengths due to the \lya\ forest, which makes continuum identification nearly impossible. 
Consequently, we neglect spectral regions blueward of the \lya\ when searching for metal absorption lines. 
In fact, even with a reliable continuum estimate, identifying metal lines in the forest of hydrogen absorption lines at the X-Shooter VIS resolution of $R = 8900$ remains highly challenging at these redshifts. 

After fitting the quasar continuum, we search for candidate absorption lines by identifying coincident absorption (i.e. at the same redshift) in pairs of transitions. 
Doublets are particularly useful in this context because we can search for matching absorption in the two transitions of the doublet, with an additional and powerful constraint on the lines relative strength given by atomic physics.
Therefore, we start by searching for the most commonly observed doublets: \civ, \nv, \siiv\ and \mgii\ (Table \ref{tab:absorption_lines}). 
We employ the \textsc{new\_system\_from\_likelihood} recipe in \textsc{Astrocook} to detect absorption peaks and estimate the probability that they correspond to the two components of the same doublet. The system is retained if the doublet is identified with a significance greater than 2$\sigma$. 
This automatic doublet identification is followed by a manual review of the identified systems, where we add or remove components as necessary.
For each doublet, the redshift search range is constrained at the lower end by the \lya\ emission line and at the upper end by the emission redshift of the quasar, as derived from the \mgii\ emission line (Section \ref{section-results-quasar-properties-emission-lines}). 
For \civ, that is by far the most frequent metal doublet identified in quasar spectra, the corresponding wavelength search range is denoted as \civ\ forest, in analogy with the \lya\ forest. 

After a first research of the most common doublets over their available redshift ranges, we proceed in our analysis with two approaches. 
First, we search for absorption lines associated with the existing systems, such as the \alii\,$\lambda 1670$ \AA\ or the \ciiALLOWED\,$\lambda 1334$ \AA\ lines, usually linked to the \mgii\ doublet. 
Second, we search for rarer pairs of transitions to identify new absorption systems, including strong \feii\ lines, either the \feii\,$\lambda \lambda 2344,2382$ \AA\ or the \feii$\,\lambda \lambda 2586,2600$ \AA\ pairs, and the \oii\,$\lambda 1302$--\siii\,$\lambda 1304$ \AA\ pair. 
These lines are typically associated with a \mgii\ doublet, which may not be detected if falling within masked or noisy spectral regions. 
The complete list of absorption lines considered in this work is presented in Table \ref{tab:absorption_lines}. 
As our major intent is to investigate the properties of NALs to characterize ionised quasar outflows, we do not delve into the systematic research of more exotic, low-ionisation transitions, as this would be irrelevant for our analysis.

After line identification, we proceed with line fitting. 
\textsc{Astrocook} fits absorption lines using Voigt profiles, where the free parameters are the system absorption redshift $z_{abs}$, the ion column density $\log (N)$ and the Doppler parameter $b$.
We impose a minimum $b$ value of $7$ km s$^{-1}$, which corresponds to $\sim 1 / 3$ of the $b$ value associated with the resolution element and defined as $c / 2R\sqrt{\ln 2}$, which assumes a value of $ \sim 22$ km s$^{-1}$ when considering a resolution of $R=8100$ typical of the NIR part of the spectrum. 

\subsection{Construction of the final absorber sample}
\label{section-x-shooter-construction-absorber-sample}

In the previous Section we described the procedure for identifying and fitting NALs across the whole redshift range available for each specific transition. 
We now focus on the associated high-ionisation absorbers that lie within $10,000$ km s$^{-1}$ of the quasar systemic redshift $z_{em}$, that is, in the quasar proximity zone. 
These systems are thought to trace quasar environments and gaseous structures that are physically bound to the central AGN, in particular ionised outflows and the host galaxy ISM and CGM. 
From now on, when referring to `systems’ and `absorbers', we implicitly mean associated systems and absorbers unless otherwise stated.

The velocity of an absorber at redshift $z_{abs}$ with respect to $z_{em}$ is estimated as $v_{abs} = \beta c$, with $c$ being the speed of light and $\beta$ computed through the relativistic Doppler formula (\citealp{perrotta_nature_2016} and references therein): 
\begin{equation}
    \beta = \frac{(1+z_{em})^2 - (1+z_{abs})^2}{(1+z_{em})^2 + (1+z_{abs})^2}.
\end{equation}
In order to construct a significant statistics, we require a $3 \sigma$ detection of the \civ$\lambda 1548$ line, the stronger component of the \civ\ doublet.
This condition translates into a threshold in the \civ\ equivalent width through the following equations  \citep{herbert-fort_metal-strong_2006}: 
\begin{equation}
    \text{S/N} \simeq \frac{3 \ \lambda_X}{c \ \omega_X} \sqrt{4.24264 \ b \ \Delta v},
    \label{eq:equivalent_width_threshold}
\end{equation}
where the S/N is the median S/N per pixel evaluated across the whole \civ\ forest extension for each quasar; $\lambda_X$ and $\omega_X$ are the rest-frame wavelength and equivalent width of transition $X$, respectively; $b$ is the Doppler parameter of the lines assumed to be $\sim 10$ km s$^{-1}$, and $\Delta v$ is the velocity resolution element per pixel, equal to 10 km s$^{-1}$. 
The threshold in equivalent width can be turned into a threshold in column density by assuming the linear regime of the curve of growth (\citealp{dodorico_metals_2016} and references therein):
\begin{equation}
    N(X) = 1.13 \cdot 10^{20} \omega_X / (f_X \lambda_X^2),
    \label{eq:column_density_threshold}
\end{equation}
where $f_X$ is the oscillator strength of transition $X$. 
Hence, by exploiting Eqs. \ref{eq:equivalent_width_threshold} and \ref{eq:column_density_threshold}, we can estimate the minimum \civ\ column density necessary to detect the \civ\,$\lambda 1548$ line at $3 \sigma$ confidence at a given S/N. 
The corresponding \civ\ rest-frame equivalent width and column density limit distributions range from 0.009 to 0.053 \AA\ and from $10^{12.4}$ to $10^{13.1}$ cm$^{-2}$, with median values of $\sim $0.024 \AA\ and $\sim 10^{12.8}$ cm$^{-2}$, respectively. 
All the identified \civ\ doublets but one satisfy the condition $ N_\civ \geq  N_{\civ, lim}$, for a total of 75 \civ\ associated absorbers distributed over 28 sources. 

Regarding instead \nv\ absorption, we identified 39 \nv\ absorbers distributed over 14 sources, with velocities up to $\sim 5400$ km s$^{-1}$.
Due to the proximity of \nv\ lines to the \lya, we can detect \nv\ systems up to a maximum velocity separation of $\sim 5600$ km s$^{-1}$ from $z_{em}$ before they fall into the \lya\ forest. 
However, \cite{perrotta_hunting_2018} showed that no substantial \nv\ absorption is found at velocities beyond $\sim 5000$ km s$^{-1}$, even when considering stacked spectra to smooth the contamination effects of the \lya\ forest. 
We confirm this by thoroughly searching for \nv\ absorption associated to \civ\ at velocities between 5400 and 10,000 km s$^{-1}$ through visual inspection, finding none. 

Lastly, we group individual absorption components closer than 200 km s$^{-1}$ into single absorption systems, whose total velocity is given by the column density-weighted average of the single absorber velocities, and whose total column density is given by the sum of the single column densities. 
The grouping procedure is applied, for each quasar, to all transitions of all ions.
The velocity threshold that defines an absorption systems is usually set at a few hundred kilometers per second (e.g. 300 km s$^{-1}$ in \citealp{perrotta_nature_2016} for the XQ-100 sample, 200 km s$^{-1}$ in \citealp{misawa_census_2007, culliton_probing_2019} and in \citealp{davies_xqr-30_2023} for the XQR-30 sample), and it represents the maximum velocity at which two absorbers are assumed to be physically bound or, in other words, part of the same physical system. 
In this work, we have assumed a velocity threshold of 200 km s$^{-1}$, following a procedure similar to that outlined in \cite{davies_xqr-30_2023}.
We examined the distribution of velocity separation between each absorption component and the nearest neighbouring component of the same transition and found that the cumulative fraction steeply increases up to the value of 200 km s$^{-1}$, then it reaches a plateau. This behaviour suggests that the majority of physically associated components have velocity separations less than 200 km s$^{-1}$. 

The clustering procedure leads to 50 \civ\ associated absorption systems and 22 \nv\ associated absorption systems, for a total of 19 associated systems presenting both \civ\ and \nv\ absorption (see Table \ref{tab:intrinsic_systems_summary} and Section \ref{section-results-statistics} for more details).  

\subsection{Modelling the quasar UV emission spectrum}
\label{section-x-shooter-modeling-uv-spectra}

\begin{figure}
    \centering
    \includegraphics[width=\linewidth]{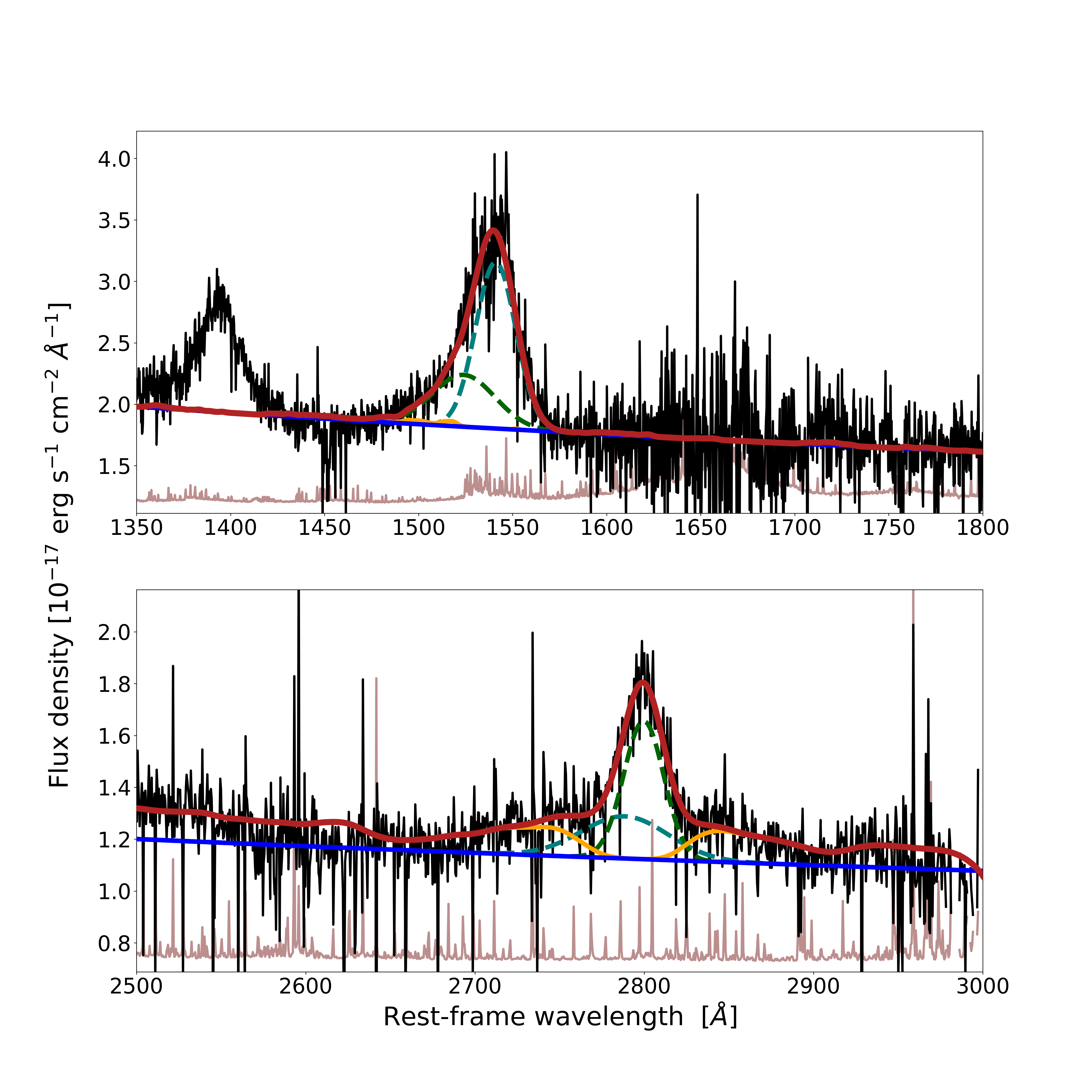}
    \caption{Best-fit decomposition of \civ\ (upper panel) and \mgii\ (lower panel) spectral regions for J2226-0618. 
    The blue line is the power law continuum, the orange line is the \feii\ template, green dashed lines show Gaussian components. The \siiv\ emission line at $\sim 1400$ \AA\ is masked during the fitting procedure.}
    \label{fig:mgii-civ-fits}
\end{figure}

In this Section, we focus on the fit of \civ\ (1549.48 \AA) and \mgii\ (2799.117 \AA) emission lines and UV quasar continuum emission. 
We verify the flux calibration against \textit{i} and \textit{z} photometry from Pan-STARRS DR1 \citep{chambers_pan-starrs1_2016}, and perform the absolute flux calibration by normalising the total spectrum to the \textit{z}-band photometry. 
For those objects displaying BAL troughs, we consider the SDSS-based empirical templates (Sect. \ref{section-x-shooter-BALs}) as reference quasar continuum for the photometric normalisation.
We attribute potential discrepancies between spectroscopic and photometric data primarily to slit losses rather than intrinsic quasar variability over different times of observations. 
This assumption is supported by previous works in the XQR-30 sample \citep{bischetti_suppression_2022, bischetti_fraction_2023}, which found no significant variability in quasar colours over at least five years of photometric observations, with magnitudes measured at different epochs consistent within statistical photometric uncertainties. 
The normalisation factors that we derive correspond to luminosity variations of less than a factor of two, translating into black hole mass variations of $\sim 0.1$ dex -- well within the statistical uncertainties on mass measurements, and significantly smaller than the systematic uncertainties of $\sim 0.3$ dex associated with virial relations (Sect. \ref{section-results-uv_analysis}).
These results validate our approach of normalising the spectra to the \textit{z}-band, ensuring no  introduction of bias or additional uncertainties.

Subsequently, we use \textsc{Astrocook} to re-sample the X-Shooter spectra into 50 km s$^{-1}$-width bins to increase the S/N.
We construct a physically motivated model for the quasar UV emission spectrum, comprising a power law combined with a \feii\ pseudo-continuum for the continuum, and multiple Gaussian components for the emission lines.
The adopted \feii\ pseudo-continuum is based on the empirical template from \cite{vestergaard_empirical_2001}, and is convolved with a Gaussian kernel of variable standard deviation (500-5000 km s$^{-1}$) to account for the velocity distribution of broad line region (BLR) gas. 

We perform local fits around the \civ\ and \mgii\ emission lines.
These local fits provide reliable estimates of the line and continuum parameters without delving into the complex task of modelling the wavelength-dependent quasar reddening that becomes significant blueward of the \civ\ line and often requires a second- or third-order polynomial correction (e.g. \citealp{shen_gemini_2019, mazzucchelli_xqr-30_2023}). 
We first fit the quasar continuum in line-free spectral regions.  
We mask the noisy region of transition between the VIS and NIR arms at observed wavelengths of 1000-1020 nm, and those regions contaminated by strong telluric lines at observed wavelengths of 1340--1450 nm and $\gtrsim 1800$ nm. 
After the subtraction of the best-fit continuum model from the data, we fit the residual emission lines with one to three Gaussian components. We then select the line model with the lowest reduced $\chi ^2$ value. 
The \civ\ emission line is best-fitted with one, two, and three Gaussian components in 11, 27, and 1 quasar(s), respectively. 
For the four quasars in our sample with $z_{em} \gtrsim 5.4$, the \mgii\ line falls in the telluric contaminated region. For two of them (J0108+0711 and J2207-0416), we still managed to fit the \mgii\ line; however, for the other two (J1335+0328 and J1022+2252) we could not disentangle \mgii\ from telluric emission. 
The \mgii\ line is thus modelled using one and two Gaussian components in 3 and 34 quasars, respectively, and not detected in two quasars.
An example of line best fit is shown in Fig. \ref{fig:mgii-civ-fits}. 
We report the line and continuum parameters of the best fit solutions in  Table \ref{tab:quasar_properties}, discussed in detail in Section \ref{section-results-quasar-properties-emission-lines}.

Statistical uncertainties on the continuum and line parameters are evaluated as follows. 
For each source, we generate N=1000 realisations of the \civ\ and \mgii\ line profiles and underlying continua, assuming that the flux in each spectral channel follows a Gaussian distribution with standard deviation equal to the associated spectral error.
For each realisation, we perform the spectral fits as described above, and adopt the median and standard deviation of the resulting continuum and line parameter distributions as best-fit values and associated errors, respectively. 
We find that the parameter with the largest uncertainty is the continuum slope. This is expected, given that we are performing local fits over relatively narrow spectral regions of $\sim 150$\AA.

\subsection{Identification of BALs}
\label{section-x-shooter-BALs}

\begin{table}[]
    \centering
    \caption{Results of BAL analysis.}
    \adjustbox{max width=0.45\textwidth}{
    \begin{tabular}{ccccc}
    \toprule
       Quasar & BI$_\text{\civ}$ & $v_{min, \text{\civ}}$ & $v_{max, \text{\civ}}$ & BI$_\text{\siiv}$ \\
        & [km s$^{-1}$] & [km s$^{-1}$] & [km s$^{-1}$] & [km s$^{-1}$]\\
    \midrule
       J0017-1000 & $4430^{+1380}_{-580}$ & $9570 ^{+2700}_{-3500}$ & $21140^{+1850}_{-2330}$  & / \\[5pt]
       J0338+0021 & $890^{+540}_{-480}$ & $12410^{+290}_{-100}$ & $22490^{+180}_{-460}$  & / \\[5pt]
       J1601-1828 & $2610^{+1100}_{-820}$ & $15220^{+3820}_{-2290}$ & $40980^{+50}_{-170}$  & $1410^{+90}_{-250}$ \\[5pt]
       J1004+2025 & $6080^{+1310}_{-1290}$ & $24870^{+140}_{-230}$ & $48670^{+130}_{-90}$  & $2600^{+360}_{-460}$ \\[5pt]
       J0957-1016 & $5680^{+1140}_{-1090}$ & $9120^{+2750}_{-6020}$ & $23120^{+1100}_{-690}$ & $1280^{+1230}_{-710}$ \\[5pt]
       J2225+0330 & $5390^{+30}_{-130}$ & $9460^{+100}_{-80}$ & $23580^{+60}_{-50}$  & $3240^{+100}_{-420}$\\[5pt]
    \bottomrule
    \end{tabular}}
    \parbox{0.45\textwidth}{
    \vspace{2mm}
    \footnotesize
    \textbf{Notes:} 
    From left to right: \civ\ balnicity index, minimum and maximum outflow velocities for the six quasars in our sample displaying BAL features. The last column reports the balnicity index for the four targets showing \siiv\ BAL troughs.  
  }
    \label{tab:BAL-properties}
\end{table}

\begin{figure*}
    \centering
    \includegraphics[width=1\linewidth]{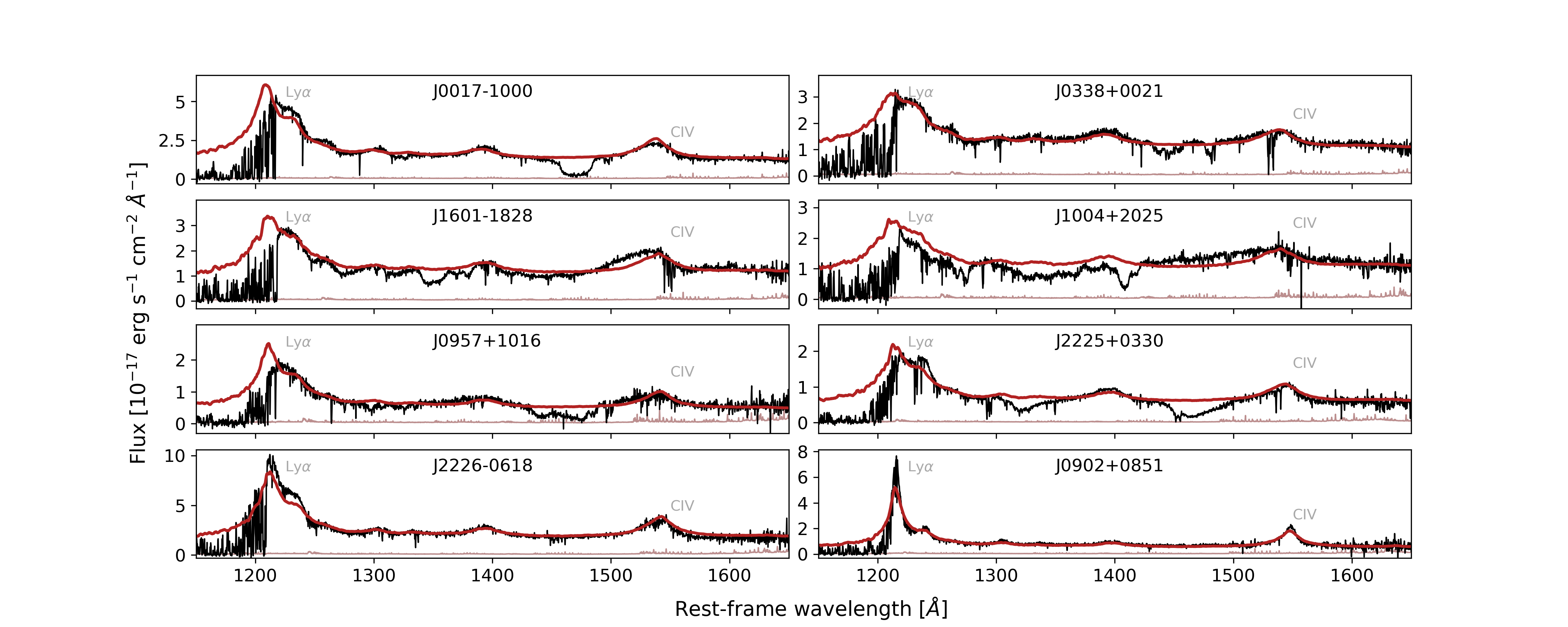}
    \caption{Spectra of the six BAL quasars in our sample and two non-BAL quasars (bottom panels) reported for comparison. Quasar spectra, binned to 50 km s$^{-1}$, are displayed in black and the associated errors in pink, while the composite SDSS templates are shown in red.}
    \label{fig:bal_sdss_templates}
\end{figure*}

We additionally search the X-Shooter spectra for BAL absorption troughs. 
The first step is to model the UV intrinsic emission of the quasar. 
The local, analytic fits performed in Section \ref{section-x-shooter-modeling-uv-spectra} do not cover a sufficiently large portion of the quasar spectrum to be suitable for this purpose. 
Moreover, even if considering wider spectral regions, such fits would result in strongly model-dependent BAL characterisation. 
Therefore, we exploit an alternative approach described in \cite{bischetti_suppression_2022, bischetti_fraction_2023}, based on the use of composite templates built from observed non-BAL quasar spectra to empirically estimate the quasar UV/optical intrinsic emission.  
Here, we briefly summarise the key steps of the method, referring to the cited papers for further details. 
For each quasar of the sample, the composite template is built as the median of 100 spectra from the \cite{shen_catalog_2011} SDSS quasar catalogue, selected to match within $\pm 20$\% the original quasar colours and \civ\ equivalent width, and classified as non-BAL in \cite{gibson_catalog_2009}. 
In particular, for the quasar colours the following flux ratios are considered: $F$(1700\AA)/$F$(2100\AA) to reproduce the continuum slope redwards of \civ, and $F$(1290\AA)/$F$(1700\AA) to account for potential changes in the slope at lower wavelengths. 
The composite template is normalised to the median flux value of the quasar spectrum in the rest-frame 1650--1750\AA\ spectral interval. 

The procedure works well for most sources in our sample (Fig. \ref{fig:bal_sdss_templates}) except for a few quasars in which the \civ\ lines have extremely large velocity shifts or extremely large and/or asymmetric profiles. For example, this is the case for the two BAL quasars J1601-1828 and J1004+2025, whose \civ\ emission is characterised by a prominent blue wing which extends up to the BAL trough at velocities of $\sim 20,000$ km s$^{-1}$ (Table \ref{tab:BAL-properties}). 
However, this does not affect the BAL characterisation, as the discrepancies between the observed spectrum and the empirical template are in a region free from absorption.
Such discrepancies are due to the rarity of SDSS spectra with such prominent \civ\ wings, only observed in the most luminous and highly accreting SMBH (e.g. \citealp{vietri_wissh_2018, schneider_sloan_2010}). 
BALs are identified through the balnicity index (BI, \citealp{weymann1991}) criterion, which is a modified expression of the line equivalent width (EW): 
\begin{equation}
    \text{BI} = \int _0 ^{v_{lim}} C \bigg(1 - \frac{f(v)}{0.9} \bigg) dv,
    \label{eq-BI-definition}
\end{equation}
where $f(v)$ is the normalised spectrum, and $C$ is a constant factor equal to 1 if $f(v)<0.9$ for contiguous troughs larger than 2000 km s$^{-1}$, and 0 otherwise.
For \civ\ BALs, we impose $v_{lim} \sim 64,000$ km s$^{-1}$, implying a full coverage of the spectral region between \lya\ and \civ\ emission lines. Similarly, for \siiv\ and \nv\ BALs we impose $v_{lim} \sim 35,000$ km s$^{-1}$ and $6000$ km s$^{-1}$, respectively. 
The minimum ($v_{min}$) and maximum ($v_{max}$) BAL velocities are identified as the lowest and highest velocity for which $C=1$ in Eq. \ref{eq-BI-definition}, respectively. 
Errors on BAL parameters are determined following \cite{bischetti_suppression_2022}, and taking into account the uncertainties on both the slope and normalisation of the composite template. 
We identified six \civ\ BALs within our sample (Table \ref{tab:BAL-properties}), with BI spanning from $\sim 1200$ km s$^{-1}$ to $\sim 6200$ km s$^{-1}$, and maximum velocities reaching almost $\sim 49,000$ km s$^{-1}$ for J1004+2025.  

\begin{table*}[thb]
  \centering
   \caption{Summary of available ALMA archival observations and their general properties.}
   \vspace{2mm}
  \label{table-ALMA-generalities}
  \adjustbox{max width=1\textwidth}{
 
  \begin{tabular}{ccccccccc}
    \toprule
    QSO & Project ID & Ra & Dec & Central Freq.& Synth. Beam  & RMS (cont.) & RMS (cube) & \cii\ detection\\
    & & [hh:mm:ss] & [dd:mm:ss] & [GHz] & [arcsec$^2$] &  [mJy/beam] &  [mJy/beam] & at S/N$\geq$5 \\
    \midrule
    J0131-0321 & 2019.1.00840.S & 01:31:27.49 & -03:21:00.0 & 301.4 & 1.07x0.94 & 0.04 &  0.4 & Yes\\
    J0747+1153 & 2019.1.01721.S & 07:47:49.18	& +11:53:52.5 & 304.2 & 1.21x1.05 & 0.01 & 0.1 & No \\
    J0306+1853 & 2022.1.00662.S & 03:06:42.51 & +18:53:15.85 & 292.7 & 1.04x0.59 & 0.12 & 1.0 & Yes\\
    J1022+2252 & 2022.1.00662.S	& 10:22:10.04 & +22:52:25.3 & 287.7 & 0.85x0.74 & 0.05 & 0.4 & No\textsuperscript{\textit{(a)}}\\
    J2207-0416 &  2022.1.00662.S & 22:07:10.15 & -04:16:53.1 & 285.1 & 0.82x0.61 & 0.04 & 0.6 & Yes\\
    J1135-0328 & 2022.1.00662.S & 13:35:56.24 & -03:28:38.20 & 290.8 & 0.84x0.65 & 0.03 & 0.4 & Yes\\
\bottomrule
\end{tabular}}
\parbox{\textwidth}{
    \vspace{2mm}
    \footnotesize
    \textbf{Notes:} 
    \textit{(a)} Tentative detection at S/N$\sim$3, neglected in the following analysis.
  }
\end{table*}

\begin{figure}
    \centering
   \includegraphics[width = \linewidth]{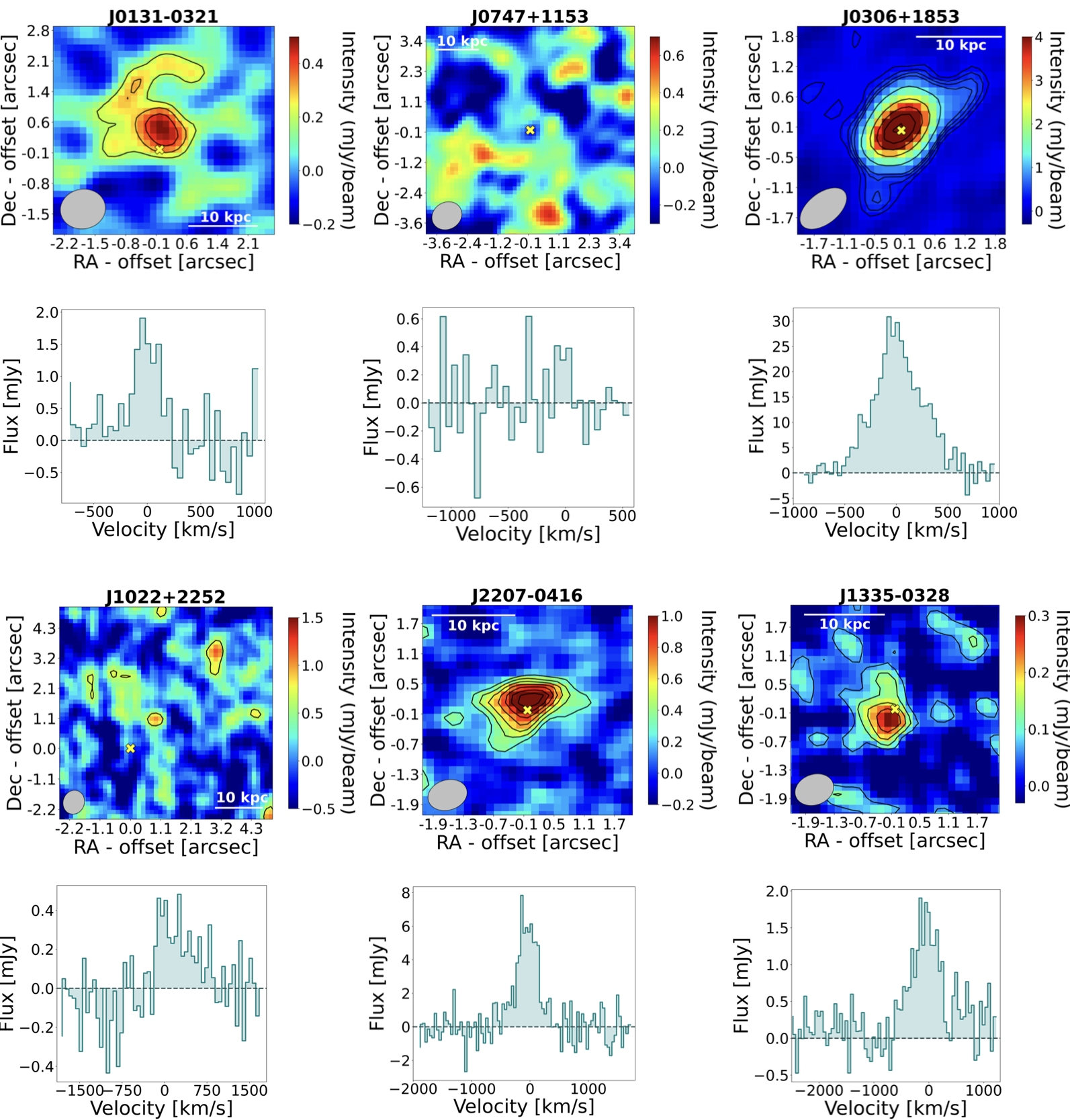}
   \caption{Two-dimensional velocity-integrated \cii\ line emission maps (upper panels) and spectral profiles (lower panels) for the six analysed sources from Table \ref{table-ALMA-generalities}. The yellow crosses in the line maps refer to the optical position of quasars, as retrieved from the SDSS \citep{wang_probing_2016}. In the same maps, we also report contour lines at 2, 3, 4$\sigma$ for J0131-0321, from 2 to 50$\sigma$ for J0306+1853, from 2 to 8$\sigma$ for J2207-0416, from 2 to 5$\sigma$ for J1335-0328 and the 2$\sigma$ contours for J1022+2252. Concerning in particular J1022+2252, we find no evidence of \cii\ emission in correspondence of the quasar optical position, although a weak signal (S/N $\sim 3$, also displayed in the correspondent lower panel) is detected with an offset of more than 2 arcsec. Instead J0747+1153 shows no evidence of \cii\ emission.}
\label{Figure-4}
\end{figure}

\section{ALMA data analysis}
\label{section-ALMA-data}

Among the 39 quasars of our sample, six have at least one ALMA archival observation covering the \cii\ line.
Here we outline the data reduction procedure applied to these data. 
We retrieve the calibrated visibilities from the ALMA science archive and perform the analysis using Common Astronomy Software Applications (\textsc{CASA}, \citealp{mcmullin_casa_2007}), version 6.6.0.20. 
We apply the \textsc{tclean} function with natural weighting, Hogbom deconvolution algorithm and a cleaning threshold of $3 \sigma$ to calibrate the data cube and to produce continuum and emission line maps. 
In particular, to obtain the continuum map we use \textsc{tclean} in multi-frequency synthesis (MFS) mode in all the line-free channels that we identify through visual inspection of the preliminary science data available on the ALMA archive. 

For the study of \cii\ line emission, we first subtract the quasar continuum from the spectral window(s) containing the line. This is done using the \textsc{uvcontsub} task, which operates in the visibility space by fitting the continuum visibilities in line-free channels using a zero-order polynomial.
We then apply \textsc{tclean} to the continuum-subtracted visibilies, using the same set of parameters as above. 
In MFS mode, this procedure generates the velocity-integrated emission line map, and in cube mode it synthesises the data cube. 
The spectral resolution of the cube is set manually in the range 40 to 60 km s$^{-1}$.
The pixel size is fixed at 0.15" for all sources, so that the minor FWHM of ALMA synthesised beam comprises at least 5 resolution elements. 
The root-mean-square (rms) noises we obtain span from 0.03 to 0.12 mJy/beam for the continuum maps, and from 0.4 to 1.0 mJy/beam for the data cubes (Table \ref{table-ALMA-cii-lines}).  

For each source, we analyse the \cii\ velocity-integrated line map and evaluate the line S/N using the emission peak and the rms in source-free regions. 
We find that four sources present a \cii\ line detection with S/N $\geq 5 \sigma$ (Table \ref{table-ALMA-cii-lines}). 
For reference, the continuum detections span from S/N $\sim3$ to $\sim 50$. 
The velocity-integrated \cii\ maps and the corresponding line spectral profiles, extracted over a region corresponding to S/N $\geq 2 \sigma$, are reported in Fig. \ref{Figure-4}.
J0757+1153 shows a $3\sigma$ upper limit of 1.2 mJy/beam. 
For J1022+2252, the line is tentatively detected with a very low S/N ($\sim 3$) and a significant offset with respect to the quasar optical position. Hence, we excluded it from further analysis.

\begin{figure*}
   \centering
   \includegraphics[width = 0.9\linewidth]{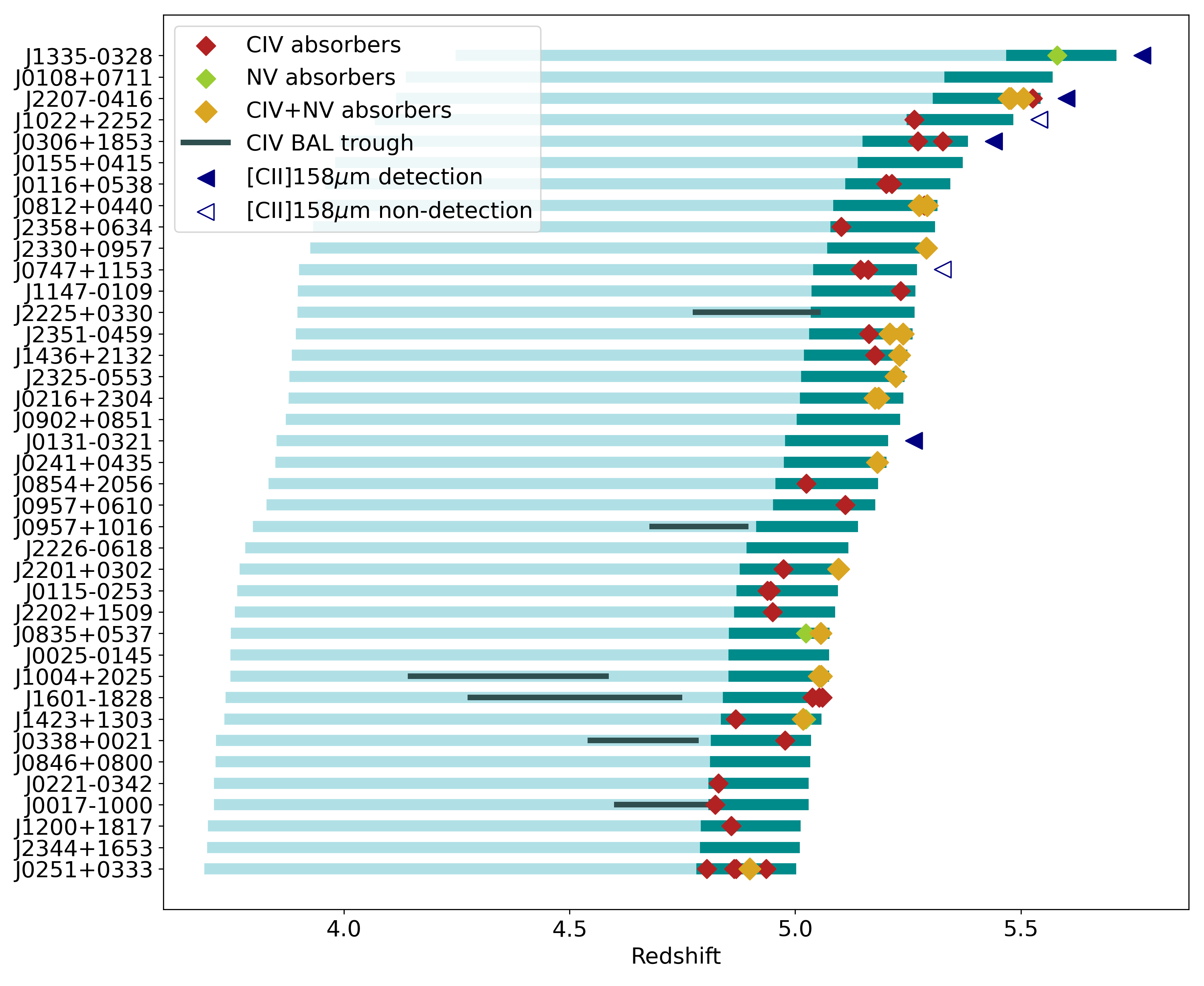}
   \caption{For each quasar of the sample, the \civ\ forest (light blue) and proximity zone (dark blue) are shown. Intrinsic absorbers are reported in different colours according to the different transitions. In particular, we highlight \civ+\nv\ absorbers in yellow, \civ\ absorbers in red and \nv\ absorbers in green. Blue filled triangles identify those quasars with ALMA observations and \cii\ detection at S/N $\geq 5$, while empty triangles refer to \cii\ non-detections (see Section \ref{section-ALMA-data} for more details).}
  \label{Figure-civ-forest}
\end{figure*}

\section{Statistics of associated absorption systems}
\label{section-results-statistics}

\begin{table}[]
    \centering
    \caption{Overview of our sample of associated, high-ionisation absorption systems.  
    }
    \adjustbox{max width=0.45\textwidth}{
    \begin{tabular}{ c c c c | c}
        \toprule
       Transitions  & $v_{abs}  \leq 2.5$  &  $2.5 < v_{abs} \leq 5.5$ & $ 5.5 < v_{abs} < 10$ & \textbf{Tot.} \\
       \midrule
         \civ & 25 & 9 & 16 & \textbf{50}\\
         \nv  & 19 & 3 & 0 & \textbf{22}\\
        \civ+\nv & 17 (68\%) & 2 (22\%) & 0 & \textbf{19 (38\%)}\\
        \civ+\siiv & 5 & 5 &  9 & \textbf{19 (38\%)}\\
       \civ+\siiv+\nv & 3 & 0 & 0 & \textbf{3 (6\%)}\\
       \civ+\mgii & 1 & 4 & 3 & \textbf{8 (15\%)}\\
       \civ+\siiv+\mgii & 1 & 3 & 2 & \textbf{6 (12\%)}\\
       \bottomrule
    \end{tabular}}
    \parbox{0.5\textwidth}{
    \vspace{2mm}
    \footnotesize
    \textbf{Notes:} 
    $v_{abs}$ is expressed in units of $10^3$ km s$^{-1}$. 
    The percentages in parentheses are evaluated with respect to the total \civ\ sample in the correspondent velocity ranges. 
  }
    \label{tab:intrinsic_systems_summary}
\end{table}

\begin{figure}
    \centering
    \includegraphics[width=1\linewidth]{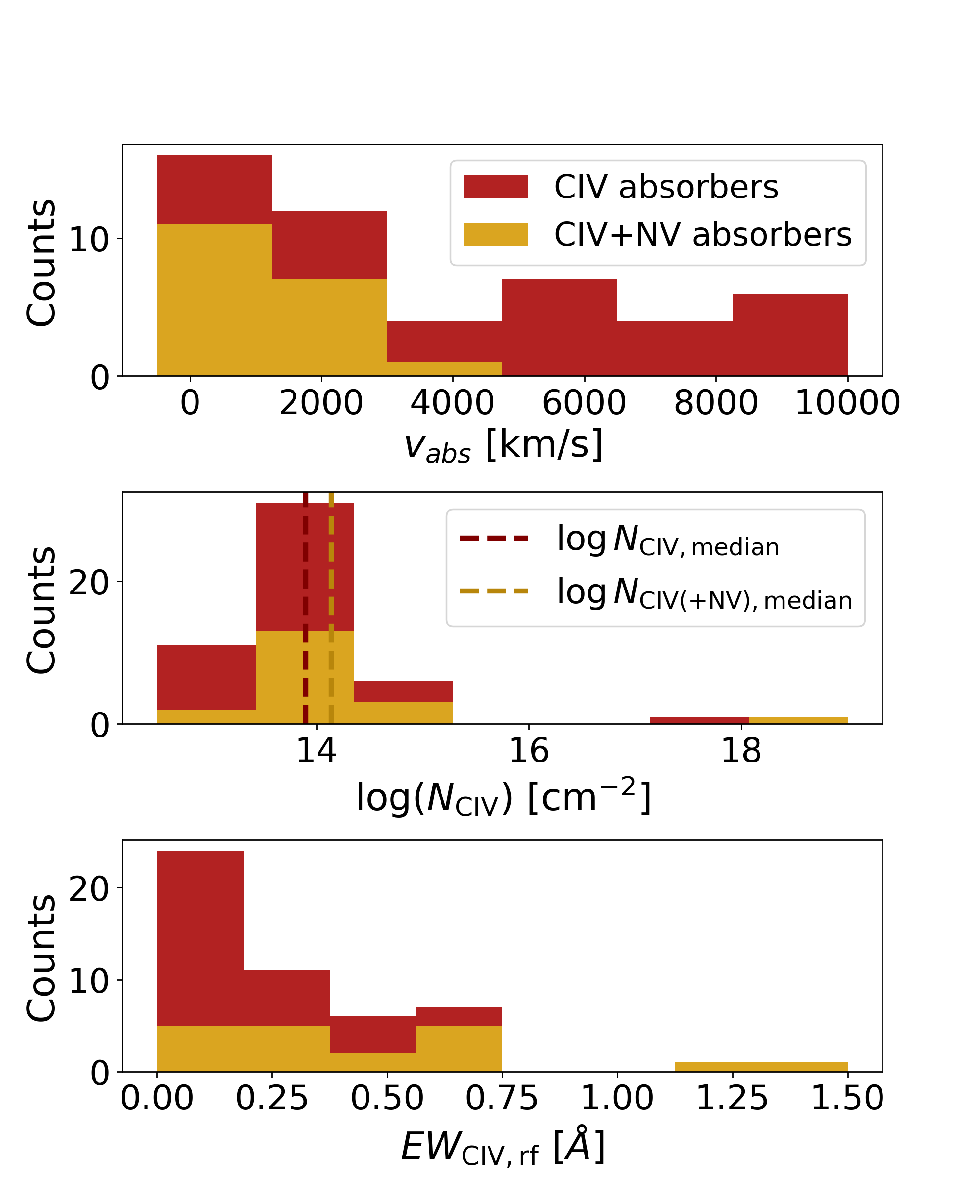}
    \caption{Properties of \civ\ (red histograms) and \civ+\nv\ (golden histograms) absorption systems. From top to bottom, distribution of absorber velocities, \civ\ column densities, \civ\ equivalent widths.}
    \label{fig:hist_associated_systems_properties}
\end{figure}

\begin{figure}
    \centering
    \includegraphics[width=0.9\linewidth]{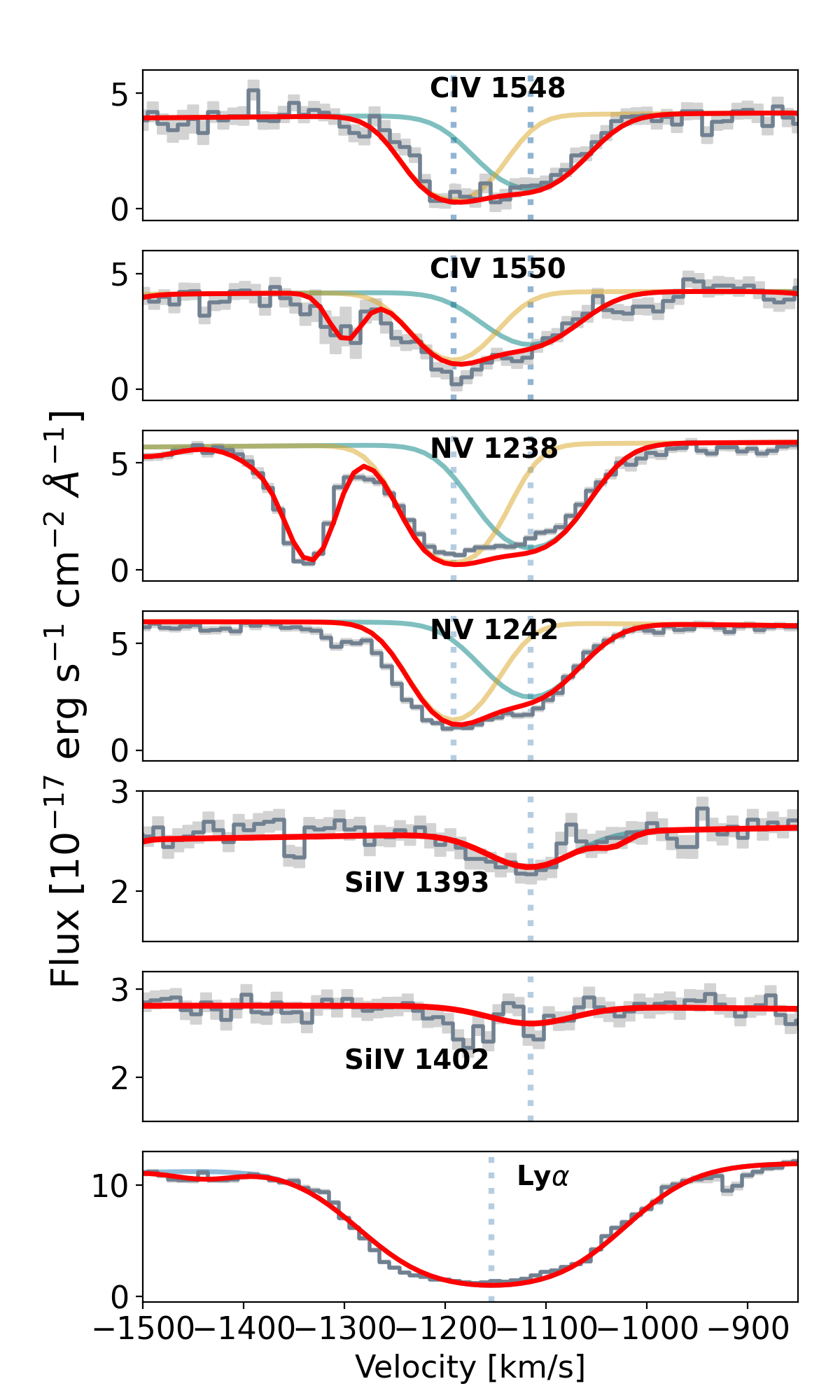}
    \caption{Associated absorption system of \civ+\nv+\siiv\ along the line of sight to J2007-0416 at $z_{abs} = 5.506$, corresponding to $v_{abs} = -1166$ km s$^{-1}$. Spectral errors are reported as a gray-shaded region.
    The associated \lya\ absorption line is also displayed. Higher-order Lyman lines are not reported as they are heavily blended with other hydrogen lines; hence they do not provide any additional information on the absorption system. 
    This system is particularly interesting, as it displays partial coverage effects on both \civ\ and \nv\ absorption systems.}
    \label{fig:J2207-0416_1200kms}
\end{figure}

Absorption systems displaying both \civ\ and \nv\ doublets constitute 37\% and 86\% of \civ\ and \nv\ absorption samples, respectively. 
Only three of the 22 \nv\ systems do not show a \civ\ counterpart. 
We analysed them individually, and found that for these systems \civ\ would either fall into the noisy superposition region of VIS and NIR arms (as it happens for the \nv\ system in J1335-0328, the highest-redshift quasar of our sample), or be blended with other significantly strong components of different absorption systems (this is the case for the two \nv\ systems detected in J0835+0537 and J1423+1303). 
Hence, it is likely that these \nv\ systems actually present an associated \civ\ absorption, but we could not detect it at the low S/N of our spectra. 
Figure \ref{Figure-civ-forest} summarises our findings. 
For each quasar of the sample, the \civ\ forest is reported in light blue, with the proximity zone highlighted in darker colour. 
Both \civ\ and \civ+\nv\ associated absorption systems are reported as red and golden diamonds, respectively. 
Their velocity, column density and equivalent width distributions are further displayed in Fig. \ref{fig:hist_associated_systems_properties}, while individual values are listed in Table \ref{table_absorption_systs_properties}.
For completeness, we also report in Fig. \ref{Figure-civ-forest} the three \nv\ systems with no \civ\ counterpart, and mark with a blue (empty) triangle those quasars with ALMA (non-)detections of \cii\ in emission.

The fraction of quasars that host at least one \civ+\nv\ absorption system is $33$\% (13 sources over 39).
The median \civ\ absorption system column density is $\log N_{\civ, median} = 13.90$ cm$^{-2}$, which increases to $14.14$ cm$^{-2}$ if we consider only \civ\ systems with associated \nv\ (Fig. \ref{fig:hist_associated_systems_properties}, middle panel), implying that \nv\ absorption is preferentially found in correspondence of strong \civ\ systems. 
The two \civ\ systems with $ \log N_{\text{\civ}} > 16$ cm$^{-2}$ display saturated lines and very poor S/N. 
Among the 50 identified \civ\ absorption systems, 25 have $v_{abs} \leq 2500$ km s$^{-1}$ and, of these, 17 (65\%) display also \nv\ absorption; 9 have 2500 km s$^{-1}$ $< v_{abs} \leq $ 5500 km s$^{-1}$, with two system showing \nv\ absorption; lastly, 16 have $v_{abs} >$ 5500 km s$^{-1}$. 
To summarise, we find 35 \civ\ absorption systems with $v_{abs} \leq 5500$ km s$^{-1}$, 19 of them ($\sim 54$\%) with associated \nv\ absorption. 
When extending the analysis to the whole proximity zone, $\sim 37$\% of \civ\ absorption systems presents a \nv\ counterpart. 
These findings are schematically displayed in Table \ref{tab:intrinsic_systems_summary}, together with additional information on other common absorption doublets, for instance \siiv\ and \mgii. 
We found 19 systems distributed over 15 sources displaying \civ+\siiv\ absorption (37\% of the \civ\ sample), and only three systems simultaneously presenting \civ, \nv\ and \siiv\ absorption (6\% of the \civ\ sample), an example of which is reported in Fig. \ref{fig:J2207-0416_1200kms}.
The velocity distribution of \civ+\siiv\ absorption systems is not peaked at small velocity separations
from $z_{em}$, confirming that the \siiv\ doublet alone is not a reliable tracer of ionised outflows.
For further comparison with previous literature absorber samples, we refer to Section \ref{section-discussion-comparison-with-previous-literature-absorption-catalogues}. 

We also investigate the occurrence of \civ\ systems with associated \mgii\ absorption. 
The simultaneous detection of these doublets, which arise from significantly different ionisation conditions, may indicate a complex environment and trace a multi-phase gaseous structure. 
We found 13 \mgii\ absorptions with $v_{abs} \leq 10,000$ km s$^{-1}$, distributed over 12 quasars. 
Among these, 8 exhibit associated \civ\ absorption, and 6 of these display also \siiv. 
We found no systems simultaneously displaying \nv\ and \mgii\ absorption, suggesting either that winds traced by \civ+\nv\ are always highly ionised, or that their multi-phase structure may be more complex than our analysis can capture. 

\begin{figure}
    \centering
    \includegraphics[width=1\linewidth]{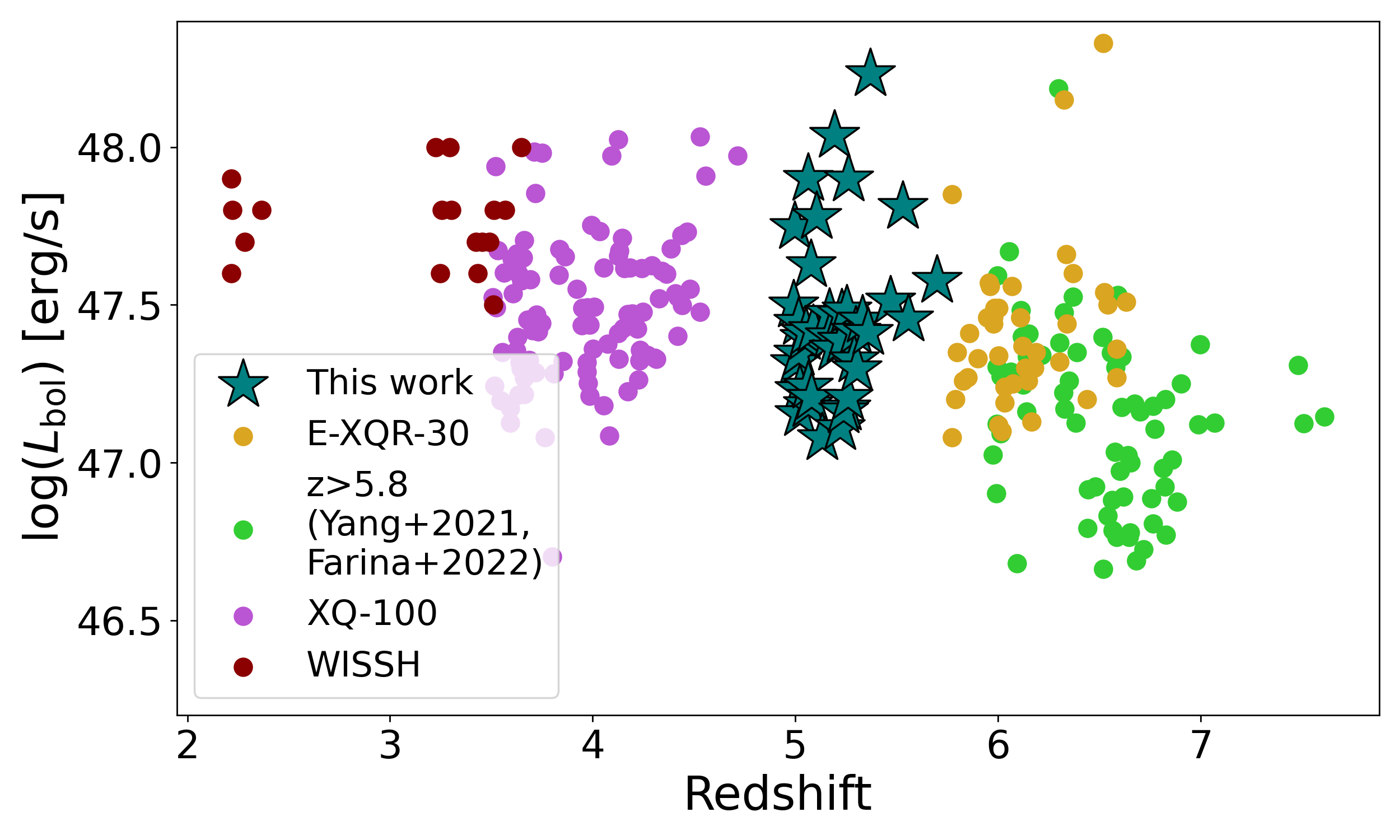}
    \caption{Bolometric luminosity versus redshift distribution for our quasar sample, reported as teal stars, flanked by other datasets from the literature covering different redshift regimes. We include the hyper-luminous WISSH quasars at $z = 2-4$ \citep{bischetti_wissh_2017, vietri_wissh_2018}, XQ-100 at $z \sim 4$ \citep{lopez_xq-100_2016}, E-XQR-30 at $z \gtrsim 6$ \citep{dodorico_xqr-30_2023}, and two $z \gtrsim 6$ datasets from \cite{yang_probing_2021} and \cite{farina_x-shooteralma_2022}. Our sample covers a unique redshift regime bridging cosmic noon and the epoch of reionisation.}
    \label{figure_Lbol_vs_z}
\end{figure}

\section{Quasar properties from emission lines}
\label{section-results-quasar-properties-emission-lines}

\subsection{Emission redshift, black hole mass, and bolometric luminosity}
\label{section-results-uv_analysis}

We exploit the spectral fits carried out in Section \ref{section-x-shooter-modeling-uv-spectra} to derive quasar redshifts, black hole masses and bolometric luminosities, reported in Table \ref{tab:quasar_properties}. 
We measure the redshift from the peak of the total \mgii\ fit profile for all quasars but  J1335+0328 and J1022+2252, in which  \mgii\ is undetected because of telluric contamination.  
For J1335-0328, we use the redshift evaluated from the \cii\ emission line, $ z_{em} = z_\text{\cii} = 5.699$ (see Section \ref{section-results-cii-emission-and-properties}). 
For J1022+2252, we adopt as reference redshift $z_{em}=5.471$, estimated in \cite{becker_evolution_2019} from the \lya\ forest. 
Our sample spans from $z \sim 5$ to $z \sim 5.7$ (Fig. \ref{figure_Lbol_vs_z}). 
We calculate the velocity shift between \civ\ and \mgii\ lines, $\Delta v$(\civ-\mgii), by measuring the difference in their peak positions.
The black hole masses are estimated from the \cite{vestergaard_mass_2009} and \cite{vestergaard_determining_2006} scaling relations for \mgii\ and \civ\ lines, respectively.
We consider the FWHM of the total line profiles, and correct the \civ\ FWHM for the presence of outflowing components using the correction from \cite{coatman_correcting_2017}.
The introduced \civ\ blueshift is defined as $c \cdot (1549.48 - \lambda_{\text{half}}[\text{\AA}])/1549.48$, with $c$ being the velocity of light in km s$^{-1}$ and $\lambda_{\text{half}}$ the wavelength that bisects the cumulative total flux of the \civ\ line. 
With this definition, positive values define blueshifts.
The \civ\ blueshift distribution spans from -150 km s$^{-1}$ to 4900 km s$^{-1}$, with median of $\sim 2100$ km s$^{-1}$. 
The frequent occurrence of prominent blueshifts and asymmetric line profiles implies that the \civ\ line is more affected by non-virial components, mostly outflows, with respect to \mgii. Additionally, the presence of strong narrow absorption systems complicates the reconstruction of the intrinsic \civ\ line profile. 
For these reasons, the \mgii\ line is considered a more reliable black hole mass estimator \citep{saturni_2018, schindler_x-shooteralma_2020}. 
From the 3000 \AA\ monochromatic luminosity we evaluate the quasar bolometric luminosity $L_{\text{bol}}$ using the bolometric correction from \cite{richards_spectral_2006}, and estimate the Eddington ratio as $\lambda _{\text{Edd}, \text{\mgii}} = L_{\text{bol}}/L_{\text{Edd},\text{\mgii}}$, where $L_{\text{Edd},\text{\mgii}}$ is the Eddington luminosity with the black hole mass estimated from \mgii.
All these quantities are reported in Table \ref{tab:quasar_properties}.

Statistical uncertainties on quasar properties are evaluated by propagating errors from line and continuum parameter uncertainties, as derived in Section \ref{section-results-uv_analysis}, without considering parameter degeneracies. 
However, this does not affect our analysis, as our focus is on global line properties (FWHM, \civ\ blueshift) derived from the total line profiles, rather than from the individual line components.
Potential deviations in the Iron emission from the \cite{vestergaard_empirical_2001} template are instead incorporated through the uncertainty in the template normalisation.

We underline that the errors reported in Table \ref{tab:quasar_properties} on black hole masses account only for the statistical error propagation, but do not consider the systematic uncertainty deriving from the single-epoch virial relations, that can be as high as $\sim$0.5 dex \citep{vestergaard_mass_2009, lai_xqz5_2024}. 
We adopt this value as systematic error on our data, and observe that the systematic uncertainty provides the major contribution to the final error on $M_{\text{BH}}$, as the statistical uncertainty is characterised by a median value of $\sim 0.1$ dex. 
Similarly, although the statistical uncertainty on $L_{\text{bol}}$ is significantly small ($\sim 0.1$ dex), the systematic uncertainty deriving from the assumption of a mean quasar spectral energy distribution can be as high as $50 \%$ \citep{richards_spectral_2006}. 
Hence, we assume a systematic error on $L_{\text{bol}}$ of $0.3$ dex, following \cite{lai_xqz5_2024}. 

\begin{figure}
   \centering
   \includegraphics[width = 0.9\linewidth]{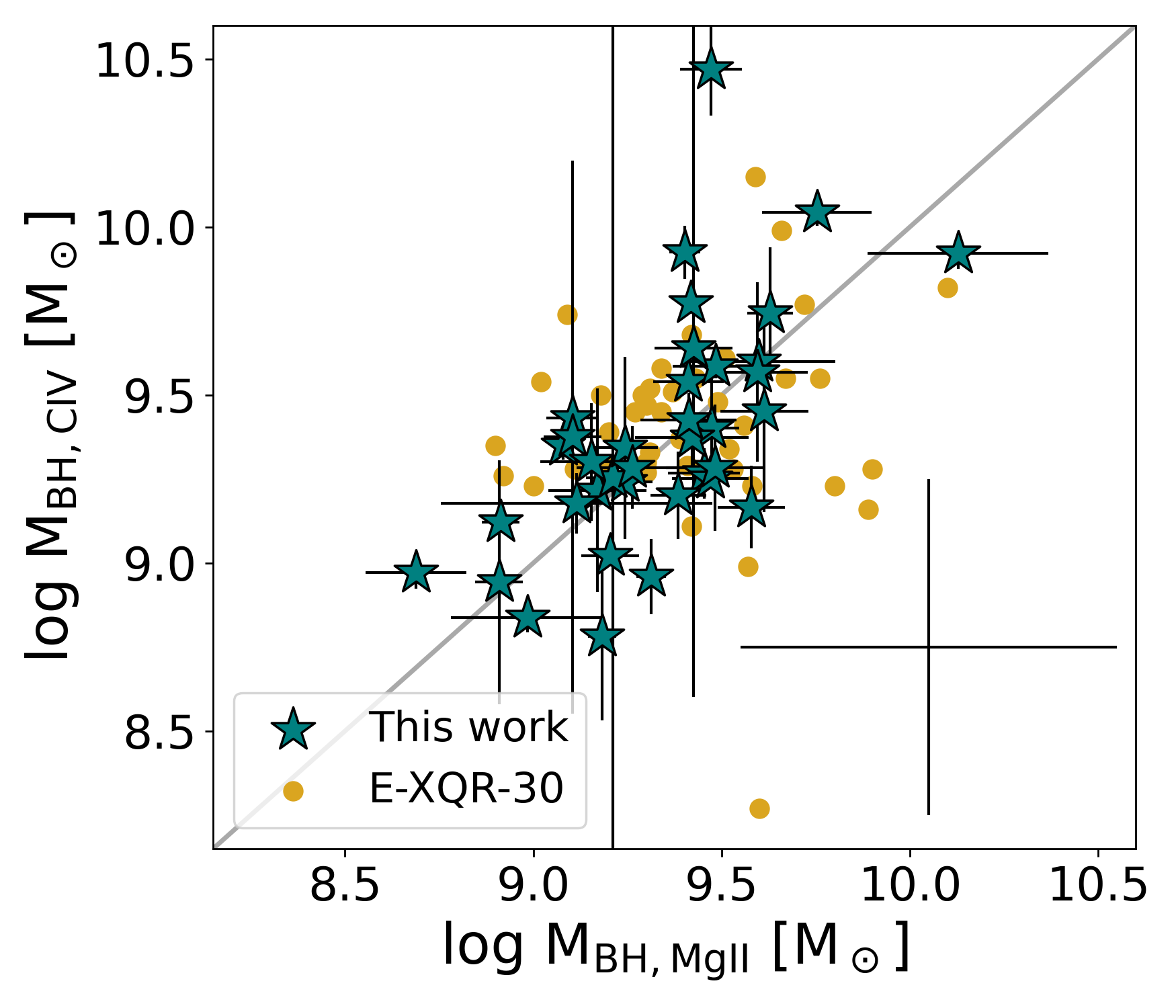}
   \caption{Comparison between \civ- and \mgii-based  black hole masses for our sample of $z\sim 5 - 5.7$ quasars (teal diamonds) and for the E-XQR30 sample at $z \gtrsim 6$ (golden data points) from \cite{mazzucchelli_xqr-30_2023}. 
   The reported errorbars account only for the statistical contribution, while the systematic uncertainty due to the scaling relations ($\sim 0.5$ dex) is reported in the right-bottom corner, and is the same for the two samples.}
\label{Figure-mbh-comparison}
\end{figure}

In Fig. \ref{Figure-mbh-comparison} we show a comparison between the \civ\ and \mgii-derived black hole masses for our sample of quasars at $z=5-5.7$, together with the Enlarged XQR-30 (E-XQR-30) sample \citep{dodorico_xqr-30_2023, mazzucchelli_xqr-30_2023}, that comprises 42 luminous quasars at  $z = 5.8 - 6.6$ observed with X-Shooter. 
We find consistent mass ranges for the two datasets: $\log M_{\text{BH}, \mgii} [M_\odot] \simeq  8.7 - 10.1 $ for our sample and $\log M_{\text{BH}, \mgii} [M_\odot] = 8.9 - 10.1$ for E-XQR-30. 
The scatter is also comparable. 
In our sample, the \civ- and \mgii-based  mass estimates are on average consistent within $\sim 0.3$ dex, exhibiting a mean $M_{\text{BH}, \text{\civ}}/M_{\text{BH}, \text{\mgii}}$ of $ \sim 1.4$, consistent with the 1.3 ratio reported in \cite{mazzucchelli_xqr-30_2023} for E-XQR-30. 
A significant outlier is J2344+1653, which reports a \civ-based mass about one order of magnitude higher than that recovered by \mgii. This is also the object in our sample with the smaller \civ\ blueshift in the line profile, roughly consistent with zero value (Table \ref{tab:quasar_properties}). These issues are likely caused by an incorrect fit of the \civ\ line for J2344+1653, due to the low S/N of the spectrum in this specific region.

\begin{figure}
   \centering
   \includegraphics[width = 1.0\linewidth]{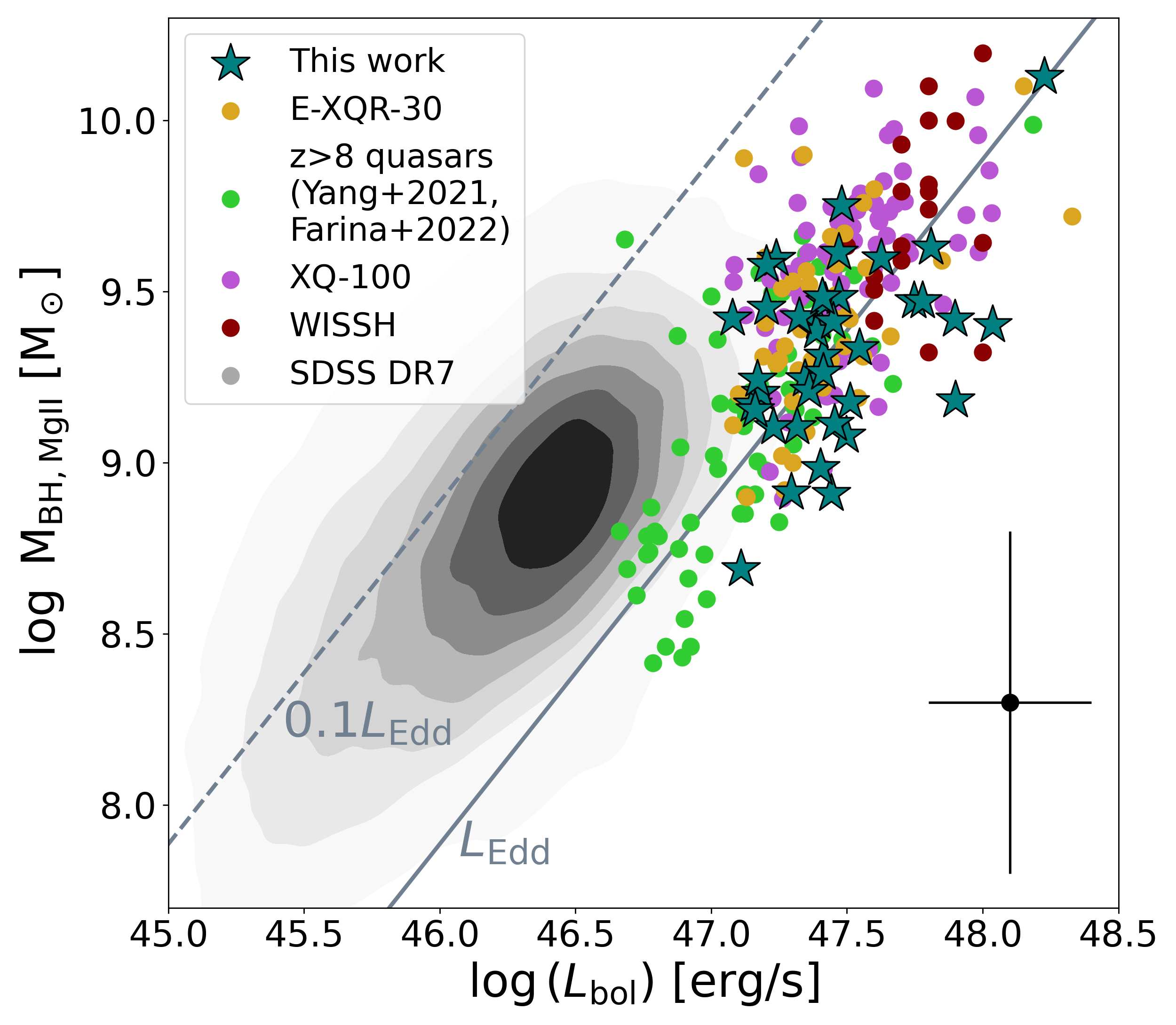}
   \caption{Black hole masses versus bolometric luminosities. The quasars analysed in this work are reported as teal stars, compared to literature quasar samples at both higher (E-XQR-30, in yellow, and compilations from \citealp{yang_probing_2021} and \citealp{farina_x-shooteralma_2022}, in green) and lower (XQ-100, in purple, and WISSH, in dark red) redshift. We note that for WISSH objects black hole masses and bolometric luminosities are estimated from the \hb\ line and not from the \mgii, as for all the other samples. Lastly, quasars at $0.35<z<2.25$ from the SDSS DR7 quasar survey \citep{shen_catalog_2011} are reported as black and grey contour lines. Typical systematic uncertainties on black hole masses ($\sim 0.5$ dex) and bolometric luminosities ($\sim 0.3$ dex) are shown in the bottom right corner.}
  \label{Figure-MBH-vs_Lbol}
\end{figure}

\begin{figure}
    \centering
    \includegraphics[width=0.95\linewidth]{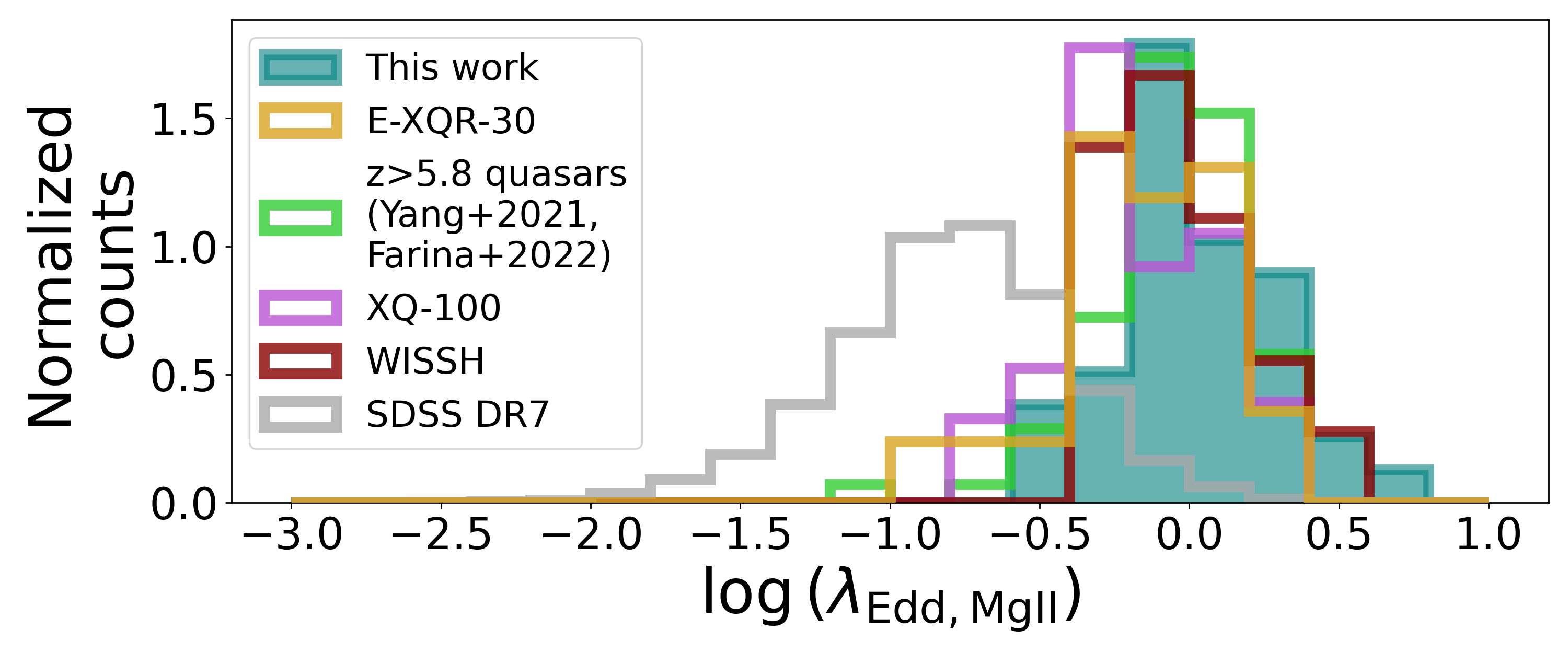}
    \caption{Distribution of the logarithmic Eddington ratio, $\log (L_{\text{bol}} / L_{\text{Edd}, \mgii} )$, for our dataset (teal filled histogram) and for the SDSS DR7, WISSH, XQ-100, E-XQR-30, \cite{yang_probing_2021}, and \cite{farina_x-shooteralma_2022} samples, reported for comparison. 
    We have not included in the histogram the two Eddington ratio estimates for J1335-0328 and J1022+2252, that exploit the \civ\ black hole mass measurements to derive $L_{\text{Edd}}$. These objects are both classified as super-Eddington (Table \ref{tab:quasar_properties}).}
    \label{fig:Ledd_histogram}
\end{figure}
In Fig. \ref{Figure-MBH-vs_Lbol} we compare $M_{\text{BH}, \mgii}$ (for all quasars but J1335-0328 and J1022+2252, for which we use $M_{\text{BH}, \civ}$) versus $L_{\text{bol}}$ for our sample, alongside a compilation of quasars from the literature at both lower and higher redshifts. 
These datasets include the SDSS DR7 quasar catalogue \citep{shen_catalog_2011} spanning from $z = 0.35$ to $z=2.25$, the 18 WISE/SDSS selected hyper-luminous (WISSH) quasars at $z = 2.3-3.5$ \citep{bischetti_wissh_2017, vietri_wissh_2018}, the XQ-100 legacy survey \citep{lopez_xq-100_2016}, E-XQR-30 and other two high-redshift samples from \cite{farina_x-shooteralma_2022} ($z = 5.8-7.5$) and \cite{yang_probing_2021} ($z = 6.3-7.6)$. 
The XQ-100 quasar masses are derived in \cite{lai_virial_2023}.
It emerges clearly from Fig. \ref{Figure-MBH-vs_Lbol} that high-redshift quasars typically exhibit higher bolometric luminosities and Eddington ratios with respect to local sources: in fact, a significant fraction of our, E-XQR-30, \cite{farina_x-shooteralma_2022}, \cite{yang_probing_2021} and also XQ-100 samples have bolometric luminosities one order of magnitude higher than the ones of SDSS sources, and display super-Eddington ratios ($\lambda_{\text{Edd}} > 1$). 
The distribution of bolometric luminosities in our sample spans from $L_{\text{bol}} \simeq 10^{47}$ to $1.7 \cdot 10^{48}$ erg/s, with median of $ 2.6 \cdot 10^{47}$ erg/s. 
The associated Eddington ratios span from 0.3 to 4, with median (mean) of $\sim$0.9 ($\sim$1.2), for our sub-sample of 37 objects with detected \mgii\ emission. 
The fraction of these quasars with $\lambda _{\text{Edd, \mgii}} > 1$ is 40\%, and seven objects display values even higher than 2 ($\sim 20$\% of the sample).
The distribution of $\lambda _{\text{Edd, \mgii}}$ is displayed in Fig. \ref{fig:Ledd_histogram}, with the other literature datasets reported for comparison. 
In general, our results are consistent with those derived for other samples of quasars at higher redshift. 
For example, \cite{yang_probing_2021} found an Eddington ratio distribution peaked at $\lambda _{\text{Edd}, median} = 0.85$ with mean $1.08$, while \cite{mazzucchelli_xqr-30_2023} found $\lambda _{\text{Edd}, median} = 0.84$ for the E-XQR-30 sample. 
If we combine our sample, XQ-100, E-XQR-30 and the two compilations from \cite{yang_probing_2021} and \cite{farina_x-shooteralma_2022}, we find that $44\%$ of these $z>3.5$ quasars exhibit $\lambda _{\text{Edd, \mgii}} >1$.

In a recent work by \cite{lai_xqz5_2024}, the authors present a new sample of 83 $z \sim 5$ quasars (XQz5) observed with both X-Shooter and the TripleSpec4.1 instrument \citep{wilson_mass_2004} on the Southern Astrophysical Research Telescope (SOAR). 
In particular, XQz5 comprises 26 quasars in common with our dataset, marked with an asterisk in Table \ref{tab:quasar_properties}. 
Given the similarity between the two analyses, we compare them and find that our black hole mass estimates are consistent with those provided in \cite{lai_xqz5_2024}, once accounted for the different adopted \mgii\ mass calibrations. 

\subsection{\texorpdfstring{\cii}{C[II]} luminosities and dynamical masses}
\label{section-results-cii-emission-and-properties}
Among the six objects with ALMA observations, only four (J0131-0321, J0306+1853, J2207-0416 and J1335-0328) show \cii\ emission with S/N $\geq 5$, for which we carry out 1D and 2D gaussian fits. 
All the derived parameters discussed in this Section are reported in Table \ref{table-ALMA-cii-lines}.
The 2D fit is performed on the velocity-integrated line map with the \textsc{imfit} task within CASA by assuming a 2D gaussian profile. 
This fit allows us to derive the angular extension and the integrated flux density of \cii\ emission, that results marginally resolved in all sources. 
The corresponding physical sizes are estimated as $1.5 \times$FWHM of the major axis  \citep{wang_star_2013} and span the range $3.5 - 7.5 $ kpc, broadly consistent with those observed in $z \gtrsim 6$ quasars ($1 - 5 $ kpc, e.g. \citealp{wang_star_2013, venemans_kiloparsec-scale_2020}).

We extract the \cii\ 1D spectrum over the spatial region at $\geq 2 \sigma$ in the velocity-integrated \cii\ maps, and fit it with one Gaussian function using the \textsc{curve\_fit} task from the python package \textsc{scipy} in all sources but J0306+1853, where two Gaussian functions are needed.
We derive the quasar systemic redshift $z_\cii$ from the peak of the line, as well as FWHM$_\cii$ and the 1D integrated line flux, $S \Delta v$. 
Our estimates of $z_\cii$ and FWHM$_\cii$ are consistent within $2 \sigma$ with those reported by \cite{Zhu2023} for J0306+1853, J2207-0416 and J1335-0328, and by \cite{mazzucchelli2024} for J0131-0321, based on the same ALMA observations but independently reduced and analysed. 
\cite{Zhu2023} also report a $3 \sigma$ \cii\ tentative detection for J1022+2252, consistent with our findings. 
We additionally compare our $z_\cii$ and $z_\mgii$ measurements for J0131-0321, J0306+1853 and J2207-0416, and find consistent results within $2 \sigma$ for each quasar, indicating minimal or no outflow contamination in the \mgii\ line profile.

We derive the \cii\ line luminosity, $L_{\cii}$, according to  \cite{solomon_molecular_2005}: 
\begin{equation}
    L_{\cii} = 1.04 \cdot 10^{-3} S_{\cii} \Delta v \ \nu_{\cii} (1+z)^{-1} D_L^2,
\end{equation}
where $L_{\cii}$ is measured in $L_\odot$ and $S \Delta v$ in Jy$\cdot$km s$^{-1}$; $\nu_{\cii}$ is the rest-frame frequency of the transition expressed in GHz, and $D_L $ is the luminosity distance in units of Mpc. 
We find $L_{\cii}$ of $ 10^9-10^{10} L_\odot$, in the same range with previous literature results \citep{venemans_kiloparsec-scale_2020, pensabene_alma_2020, wang_spatially_2024}.
From $L_{\cii}$ we estimate the molecular mass content by using the recent $M_{mol}-L_\text{\cii}$ relation presented in \cite{salvestrini_molecular_2024} and calibrated over $z > 6$ quasars. 
We find $M_{mol}$ spanning from $\sim 10^9 M_\odot$ to $3 \cdot 10^{10} M_\odot$.  
The errors on $M_{mol}$ reported in Table \ref{table-ALMA-cii-lines} are not taking into account the intrinsic dispersion of the relation, of $\sim0.1$ dex. 
We point out that $M_{mol}$ estimates based on the \citet{salvestrini_molecular_2024} calibration are more than one order of magnitude lower than those derived using the calibrations by \cite{zanella_c_2018} or  \cite{madden_tracing_2020}. As a matter of fact, \cite{mazzucchelli2024}, by exploiting \cite{zanella_c_2018}, estimate a molecular mass of roughly $\sim 10^{10} M_\odot$ for J0131-0321, that is about 5 times higher than our estimate.

Assuming that the \cii\ emission originates from an inclined rotating disk, we calculated the dynamical masses within the emitting region as follows \citep{wang_star_2013, feruglio_dense_2018, pensabene_alma_2020}:
\begin{equation}
    M_{\text{dyn}} \sin ^2 i = 1.16 \cdot 10^5 \big( 0.75 \ FWHM_{\cii} \big) ^2 D, 
    \label{equation-Mdyn}
\end{equation}
where $i$ is the inclination angle between the gas disk and the line of sight, assumed to be $i = 46$°, and $D$ is the source size in kpc from \cii\ measurements, defined as before as 1.5 times the major axis FWHM. 
The choice of a fixed inclination angle is dictated by the fact that our sources are only marginally resolved. 
We choose as reference $i$ the median inclination derived from the sample of $z > 6 $ resolved quasar host galaxies presented in \cite{wang_spatially_2024}, which is consistent with other recent resolved studies of high-redshift quasar hosts' (e.g. \citealp{pensabene_alma_2020}). 
Previous works based on fewer and marginally resolved sources provide similar results, $i \sim 55$° \citep{wang_star_2013}. 

Errors on the line FWHM and source size spans from $10\%$ to $30\%$.
We find $M_{\text{dyn}} \sin ^2 i$ varying in the range $(4.1 - 18.4) \cdot 10 ^{10} M_\odot$, with statistical errors of $10\%$ to $60 \%$. 
Because our targets are marginally resolved and with a single emission line measurement of moderate significance, $M_{\text{dyn}}$ estimates are strongly affected by the assumption of a fixed inclination angle and by non-circular beam shapes. 
For this reason, we conservatively attribute to our $M_{\text{dyn}}$ measurements a typical systematic uncertainty of 0.5 dex \citep{decarli_alma_2018, Bischetti2021}.
Using the same procedure, including the assumption of $i = 46$°, \cite{mazzucchelli2024} find a dynamical mass estimate of $(48 \pm 40) \cdot 10^9 M_\odot $ for J0131-0321, consistent with our results.

\begin{table*}
  \centering
   \caption{Results of \cii\ emission line analysis.}
   \vspace{1mm}
  \label{table-ALMA-cii-lines}
  \adjustbox{max width=1.0\textwidth}{
 
  \begin{tabular}{cccccccccc}
    \toprule
    QSO & $z_{\cii}$ & \cii\ size & FWHM$_{\text{[CII]}}$ & Flux Density & $S_{\cii} \Delta v$  & $L_{\cii}$ & $M_{mol}$ & $M_{\text{dyn}} \sin ^ 2i$ & $M_{\text{dyn}}$\\
    & & [arcsec$^2$] & [km s$^{-1}$] & [mJy] & [Jy km s$^{-1}$] & [$10^9L_\odot$] & [$10^9M_\odot$] &[$10^9M_\odot$] &[$10^9M_\odot$]\\
    \midrule
    J0131-0321 & 5.193$\pm$0.017 & 1.22x0.59 & 270$\pm$51 & 0.89$\pm$0.23 & 0.40$\pm$0.11 & 0.32$\pm$0.09 & 1.8$\pm$0.4 & 41$\pm$30 & 56$\pm$45\\
    J0306+1853 & 5.3702$\pm$0.0001 & 0.61x0.45 & 551$\pm$22 & 9.10$\pm$0.38 & 15.6$\pm$1.5 & 13.0$\pm$1.2 & 29$\pm$2 & 111$\pm$17 & 238$\pm$38\\
    J2207-0416 &  5.5310$\pm$0.0004 & 1.24x0.53 & 404$\pm$31 & 3.7$\pm$0.4 & 2.7$\pm$0.3 & 2.4$\pm$0.2 & 8.1$\pm$0.6 & 123$\pm$27 & 155$\pm$31\\
    J1335-0328 & 5.699$\pm$0.001 & 0.78x0.31 & 646$\pm$60 & 0.48$\pm$0.12 & 1.05$\pm$0.12 & 0.96$\pm$0.11 & 4.0$\pm$0.4 & 184$\pm$72 & 212$\pm$87\\
\bottomrule
\end{tabular}}
\parbox{\textwidth}{
    \vspace{2mm}
    \footnotesize
    \textbf{Notes:} 
    Redshift is estimated from the line peak, \cii\ size from 2D gaussian fit, \cii\ FWHM from 1D gaussian fit, flux density integrated over the $2 \sigma$ region, integrated flux density, \cii\ luminosity, molecular and dynamical masses. Errors on masses account only for the statistic uncertainty. The systematic uncertainty on $M_{mol}$ is of about 0.1 dex, and of 0.5 dex for $M_{\text{dyn}}$.  
  }
\end{table*}

\section{Discussion}
\label{section-discussion}

\subsection{Comparison with previous literature absorption catalogues }
\label{section-discussion-comparison-with-previous-literature-absorption-catalogues}

In this Section, we compare our sample of associated absorbers with previous results from the literature, in particular the works by \cite{perrotta_nature_2016, perrotta_hunting_2018} and \cite{davies_xqr-30_2023}.
In \cite{perrotta_nature_2016}, the authors analyse a sample of absorption systems extracted from the XQ-100 sample at $z \sim 3.5-4.5$. 
This analysis is further implemented in \cite{perrotta_hunting_2018} through a stacking procedure that aims at detecting \nv\ absorption even within the \lya\ forest.
In \cite{davies_xqr-30_2023}, the XQR-30 metal absorber catalogue is presented.
Our main findings can be summarised as follows:
\begin{itemize}
    \item The fraction of quasars hosting a \civ+\nv\ associated absorption system is almost the same for the three X-Shooter samples (33\% for our sample and XQ-100, 30\% for XQR-30). 
    \item The fraction of associated \civ\ systems with $v_{abs}\leq 5500$ km s$^{-1}$ showing \nv\ absorption is of $\sim 33$\% for both the XQ-100 and XQR-30 samples. \cite{perrotta_hunting_2018} found a percentage of 38\% when considering stacked spectra, meaning that the \nv\ absorption that falls into the \lya\ forest is generally negligible.
    We, on the other hand, find a higher percentage of $\sim 54$\%, likely due to the fact that our spectra have a lower S/N, preventing us from detecting the weaker \civ\ systems.
    At lower redshifts, \cite{fechner_nature_2009} analysed a sample of 19 quasar spectra at $z = 1.5 - 2.5$ observed with the Ultraviolet and Visual Echelle Spectrograph (UVES, \citealp{dekker_design_2000}) on VLT, and found that $\sim 37$\% of associated \civ\ absorption systems with $v_{abs} \leq 5000$ km s$^{-1}$ exhibit \nv\ as well. 
    \item Moving our focus onto other absorption lines than \civ\ and \nv\ doublets, we find that 38\% of our \civ\ systems present \siiv\ absorption (Sect. \ref{section-results-statistics}). \cite{perrotta_nature_2016} found the same percentage, while in XQR-30 \cite{davies_xqr-30_2023} found a \civ+\siiv\ incidence of 53\%. 
    Only 6\% of \civ\ systems simultaneously displays associated \siiv\ and \nv\ absorptions; for XQR-30, this percentage is more than double (15\%). 
\end{itemize}

\subsection{Host galaxy -- black hole (co)evolution}
\label{section-discussion-coevolution}

\begin{figure}
    \centering
   \includegraphics[width = \linewidth]{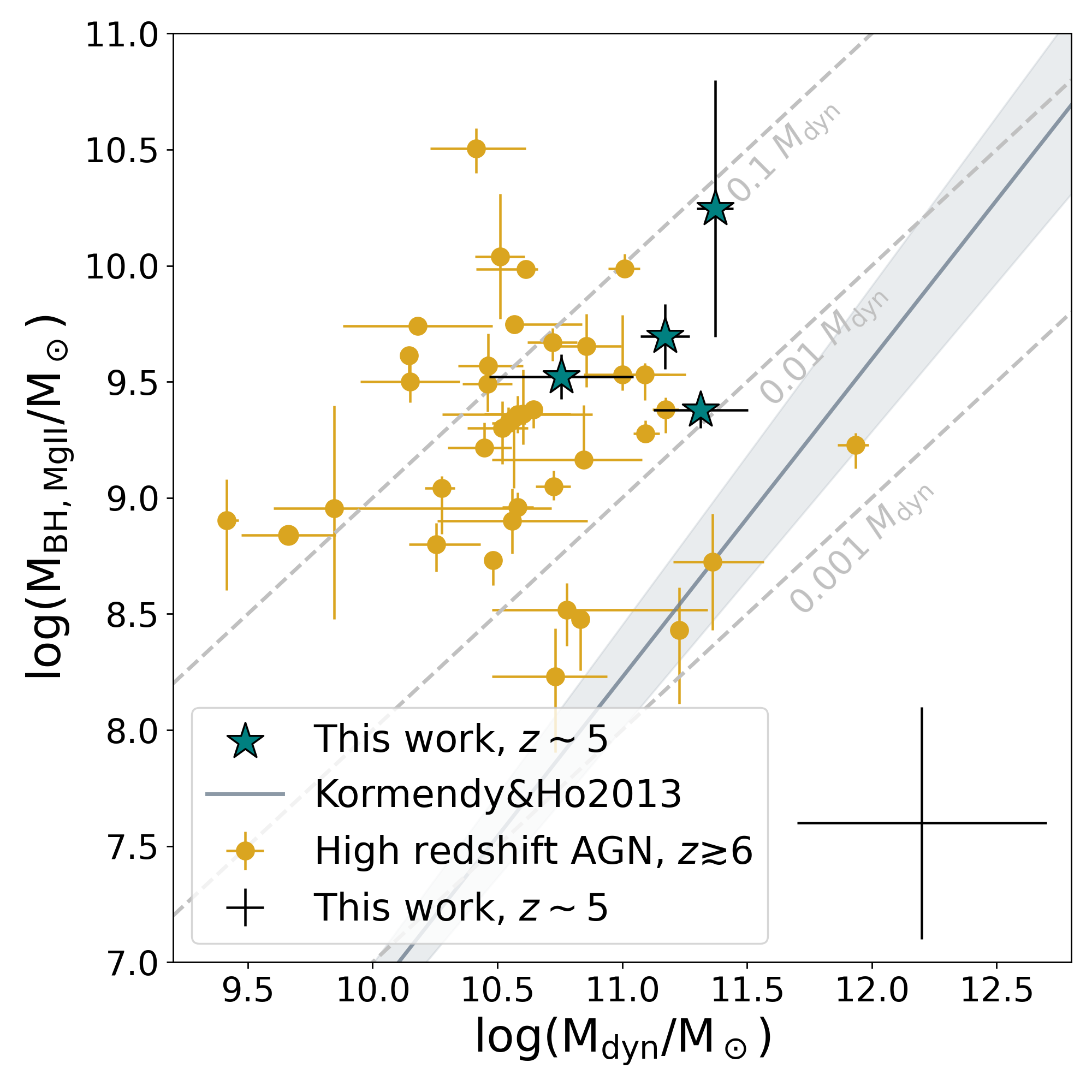}
   \caption{Black hole mass from \mgii\ versus dynamical mass for our four quasars with \cii\ S/N $\geq5$ detections (teal data points) and for other high-redshift quasars from various literature works (golden data points; \citealp{feruglio_dense_2018, pensabene_alma_2020, neeleman_kinematics_2021, tripodi_hyperion_2024, banados_far-infrared_2024,banados_blazar_2024}). 
   Systematic uncertainties of 0.5 dex on both black hole and host galaxy dynamical mass are reported in the bottom right corner. 
   The gray solid line is the local relation between black hole masses and galactic bulge masses from \cite{kormendy_coevolution_2013}, and its $1 \sigma$ uncertainty is displayed as a shaded region. }
\label{Figure-Mbh_Mdyn}
\end{figure}

We compare $M_{\text{dyn}}$ estimates with black hole masses in Fig. \ref{Figure-Mbh_Mdyn}. 
We adopt $M_{\text{BH}, \mgii}$ estimates for J0131-0321, J0306+1853 and J2207-0416, and $M_{\text{BH}, \civ}$ for J1335-0328.
We complement our four quasars with a compilation of high-redshift quasars from the literature, in particular: J2310+1855 at $z=6.00$ from \cite{feruglio_dense_2018}, for which  $M_{\text{BH}, \text{\mgii}}$ is derived in \cite{mazzucchelli_xqr-30_2023}, $z \sim 2.3 - 6.5$ quasars from \cite{pensabene_alma_2020} (see their Table 2), $ z \gtrsim 6$ quasars from \cite{neeleman_kinematics_2021} (see their Table 4), the $z \sim 7$ blazar from \cite{banados_far-infrared_2024, banados_blazar_2024}, and $z > 6$ quasars from \cite{tripodi_hyperion_2024}. 
Black hole masses are estimated from \mgii\ emission line for all literature samples. 

Our four analysed quasars lie above the local $M_{\text{BH}}-M_{\text{dyn}}$ relation, occupying in this plane the same region as $z\gtrsim 6$ quasars (e.g. \citealp{decarli_quasar_2010, decarli_alma_2018, pensabene_alma_2020, neeleman_kinematics_2021, farina_x-shooteralma_2022, tripodi_hyperion_2024, wang_spatially_2024}).
Given also their similar bolometric luminosities (Fig. \ref{Figure-MBH-vs_Lbol}), we tentatively claim that the same physical mechanisms driving quasar evolution at $z > 6$ are still in place at $z \sim 5$.
The $M_{\text{BH}}/M_{\text{dyn}}$ ratio for our four objects spans from 0.007 to 0.047. When combined with the other literature data, we find a median (mean) ratio of 0.051 (0.105), consistent with previous findings of \cite{farina_x-shooteralma_2022}.  
For comparison, in $z \sim 0$ quasars $M_{\text{BH}}/M_{\text{host}}$ is typically of $\sim 0.002$ \citep{marconi_relation_2003}, implying that the $M_{\text{BH}}/M_{\text{host}}$ ratio increases by a factor $>20$ from $z \sim 0$ to $z \gtrsim 5$. 
In previous works, \cite{decarli_quasar_2010} found an increase of a factor $\sim 7$ from $z \sim 0$ to $z \sim 3$. 
In these referred works, the host galaxy masses $M_{\text{host}}$ are estimated from photometric measurements, rather than dynamical arguments. 
More recently, \cite{wang_spatially_2024} confirmed an increase in $M_{\text{BH}}/M_{\text{dyn}}$ by a factor $\sim 20$ from $z\sim 0$ to $z >6$, although with a broad range spanning from approximately 0.6 to 60. 

A drastic decrease in $i$ by $\sim 25$° (and nearly $40$° for J0306+1853) could reconcile the deviation of our $z \sim 5$ quasars from the local relation, leading to dynamical mass estimates up to one order of magnitude higher. 
A similar argument could be extended to the other dynamical mass measurements in the literature, resulting in even more extreme variations of $i$.
However, it is unlikely that the assumption on the inclination angle alone accounts for this discrepancy, as this would imply not only that all direct $i$ measurements for spatially resolved sources are incorrect, but also that all high-redshift quasars are seen nearly face-on. 
This is statistically improbable, given the system geometry.
Thus, the most plausible explanation remains an overgrowth of SMBHs relative to their host galaxies at $z \gtrsim 5$,  though the physical mechanisms driving such rapid black hole growth remain under debate.

\subsection{Combining absorption and emission measurements}
\label{section-discussion-combining-absorption-emission}

\begin{figure*}
    \centering
   \includegraphics[width = 0.8\linewidth]{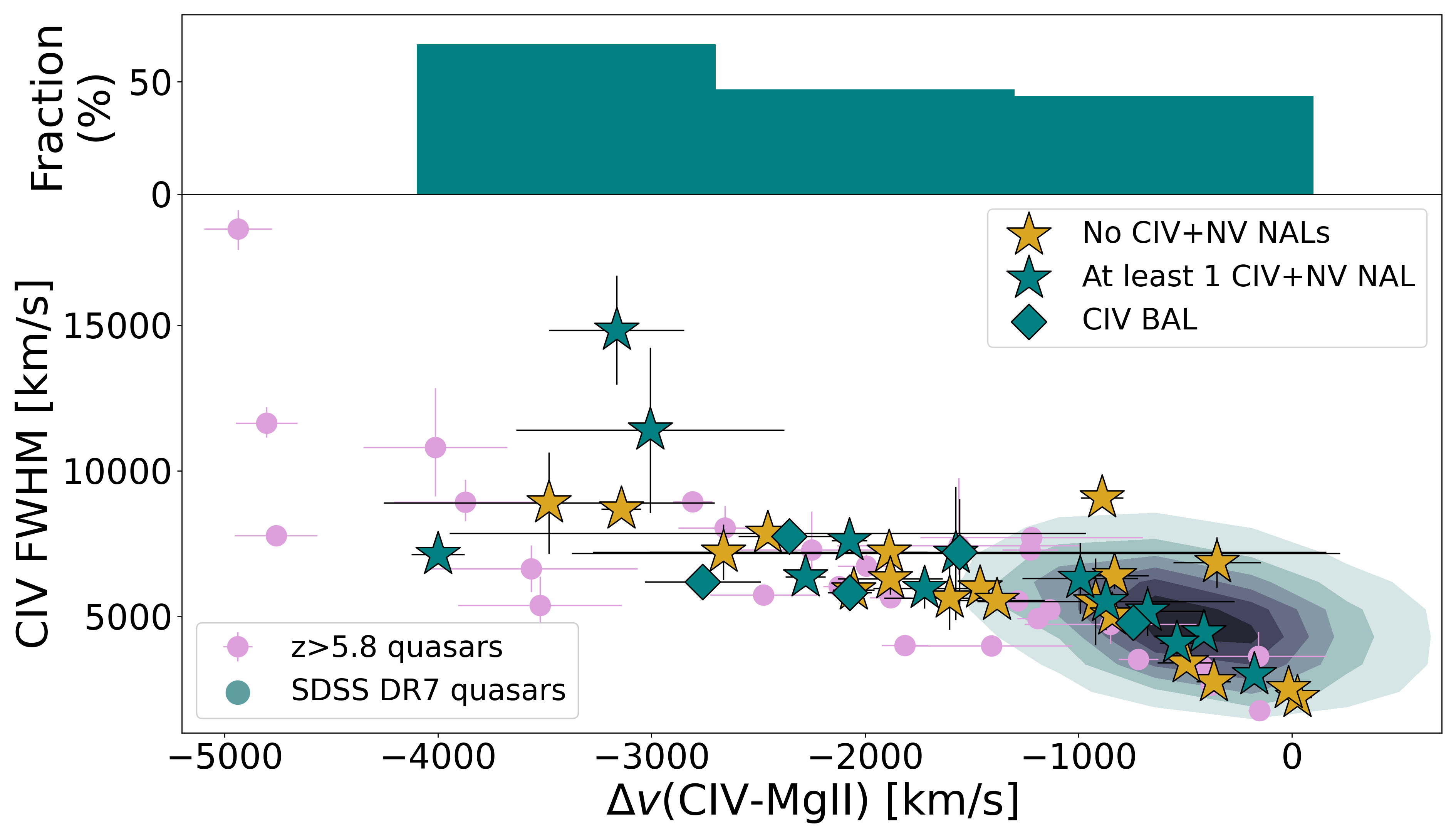}
   \caption{Comparison of \civ\ FWHM and \civ-\mgii\ velocity shift for our sample, for the $z>5.8$ quasars from \cite{schindler_x-shooteralma_2020}, and for the SDSS DR7 quasars from \cite{shen_catalog_2011}, for which the \civ-\mgii\ velocity shifts are calculated in \cite{shen_sloan_2016}. 
   The quantities on the x- and y-axis refer to emission line properties, while our dataset is colour-coded according to the presence or absence of absorption features, either \civ+\nv\ NALs (teal stars) or a \civ\ BALs (teal diamonds). 
   Quasars with a BAL trough do not typically present \civ+\nv\ NALs.
   There is only one exception in our sample displaying both, J1004+2025, for which we give priority to the NAL detection in the legend.
   The upper panel reports the fraction of quasars displaying outflow detections in absorption as a function of $\Delta v$(\civ-\mgii).}
   \label{Figure-civ_fwhm-vs-vel-shift}
\end{figure*}

\begin{figure}
    \centering
    \includegraphics[width=\linewidth]{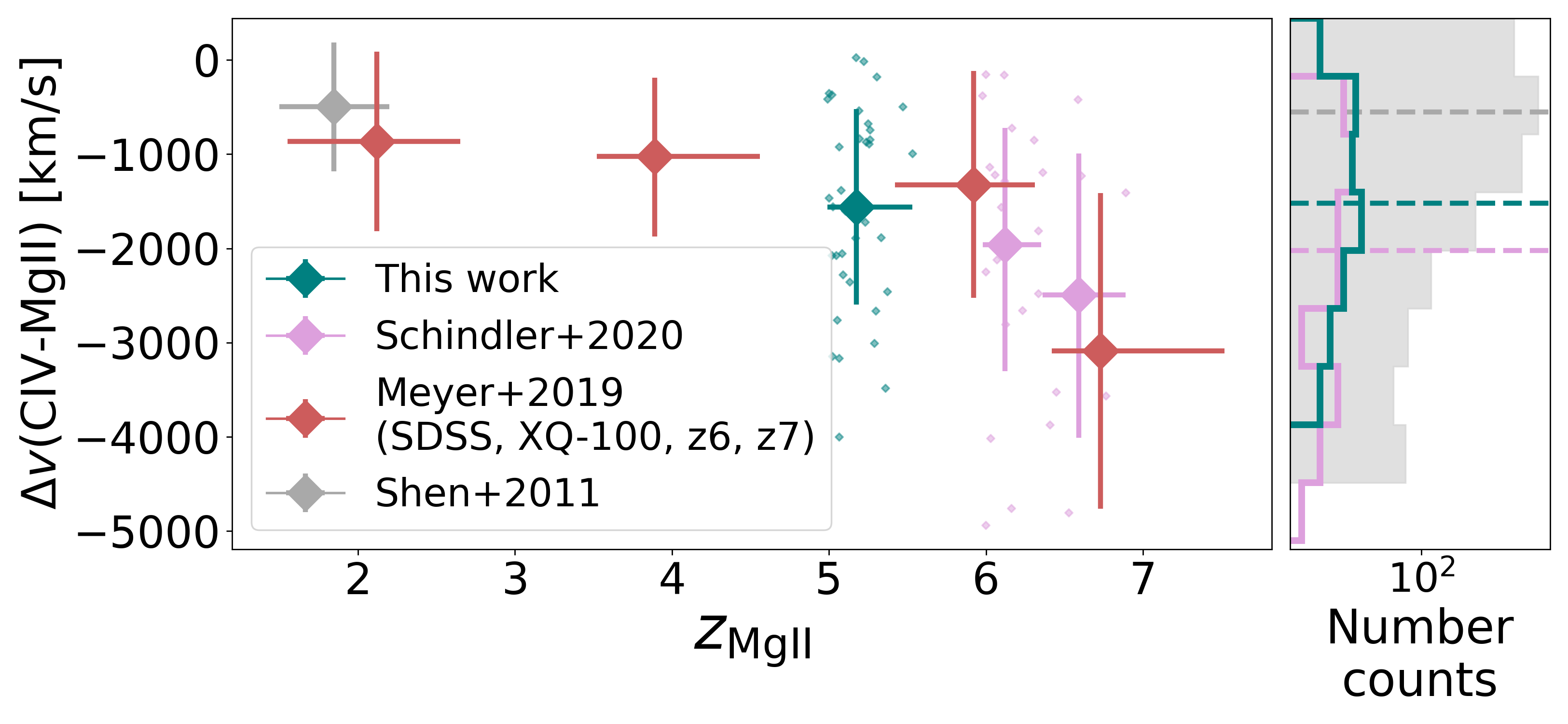}
    \caption{Mean \civ-\mgii\ velocity shift as a function of redshift from \mgii\ for our quasar sample (teal diamond), for the $z>5.8$ dataset from \cite{schindler_x-shooteralma_2020}, further divided into two redshift bins at $z \gtrless 6.35$, and for a compilation of different samples collected in \cite{meyer_new_2019}, for instance: SDSS DR7 and DR12, XQ-100 and $z \sim 6-7 $ quasars from \cite{bosman_new_2018} and \cite{schneider_sloan_2010, shen_catalog_2011}, respectively. 
    The x-axis bars mark the redshift range of each sample, while the errorbars on the y-axis report the standard deviation in each bin. }
    \label{figure-evolution-of-vel-shift-with-redshift}
\end{figure}

Studies analyzing samples of quasars at $1.5 \lesssim z \lesssim 5$ \citep[e.g.][]{misawa_census_2007,culliton_probing_2019} have shown that the fraction of intrinsic absorption systems within $5000$ km s$^{-1}$ of $z_{em}$ varies from $\sim 20$\%--$33\%$ for \civ\ systems to $\sim 38$\%--$ 75$\% for \nv\ systems.
In these works, the intrinsic nature of NALs is confirmed by the detection of partial coverage effects, which implies that the reported percentages must be considered as lower limits. 

In the following, we will assume that all our detected \civ+\nv\ NAL systems are intrinsic, specifically tracing ionised outflow. 
We notice that at least three systems in our sample of 19 \civ+\nv\ absorbers display clear signs of partial coverage (see an example in Fig. \ref{fig:J2207-0416_1200kms}). An extensive analysis of these systems will be presented in a future work. 

A comparison between the \civ\ and \mgii\ emission line profiles (Fig. \ref{fig:comparison_line_profiles}) shows that almost all quasars in our sample display clear signs of ionised outflows, probed through broad, asymmetric and blue-shifted \civ\ profiles and/or large velocity shifts between \civ\ and \mgii\ line peaks.
In order to connect these emission and absorption features typical of ionised outflows, in Figure \ref{Figure-civ_fwhm-vs-vel-shift} we compare the \civ\ FWHM and the \civ-\mgii\ velocity shift for each quasar of our sample, with the format of the data points determined by the presence in the spectrum of either a \civ+\nv\ NAL (teal star-like points) or a \civ\ BAL (teal diamond-like points). 
We note, from Fig. \ref{Figure-civ-forest}, that quasars displaying BAL troughs do not typically present \civ+\nv\ NALs, except for J1004+2025, which shows both systems.
This aligns with the general view that BALs and intrinsic NALs represent different manifestations of the same outflow phenomenon, depending on the flow geometry and orientation effects \citep{ganguly_origin_2001, nestor2008, hamann2012}. 

We observe a clear correlation between larger \civ\ FWHM and larger (in absolute value) velocity shifts, consistent with previous results \citep{schindler_x-shooteralma_2020}. 
At the same time, the fraction of quasars with detected tracers of ionised outflows in absorption increases as the velocity shift becomes more and more extreme, reaching $> 50$\% for $\Delta v$(\civ-\mgii) $\lesssim -2000$ km s$^{-1}$ (see the upper panel of Fig. \ref{Figure-civ_fwhm-vs-vel-shift}). 
This result suggests that the physical mechanisms powering ionised outflows, detected either in emission or absorption, may be related.

Previous works \citep{meyer_new_2019, schindler_x-shooteralma_2020} have shown an evolution of the \civ-\mgii\ velocity shift with redshift, with $\Delta v$ increasing from $\sim -800$ km s$^{-1}$ for quasars at $z \sim 2$ to $\sim -1950$ km s$^{-1}$ at $z \sim 6$ and $\sim -2500$ km s$^{-1}$ at $z \sim 7$. 
We find a mean $\Delta v$ of $\sim -1600$ km s$^{-1}$ for our sample of quasars at $z \sim 5$, apparently confirming this evolutionary trend (Figure \ref{figure-evolution-of-vel-shift-with-redshift}). 
However, we caution that this trend may not solely reflect a physical change in the properties of quasars with redshift. Indeed, selection biases could also play a significant role, as the vast majority of high-redshift quasars observed so far are extremely luminous and may not be representative of the entire quasar population at those early epochs, as recent results from the James Webb Space Telescope (JWST) seem to suggest \citep{Harikane2023, Maiolino2024}. 
In this context, the Extremely Large Telescope (ELT) will be fundamental to distinguish between these two scenarios, enabling high-resolution spectroscopic studies of fainter high-redshift quasars.

We investigate the existence of further correlations between the properties of quasars and those of \civ+\nv\ NAL systems, considering our sample and XQR-30. 
Previous works \citep{misawa_census_2007, stone_narrow_2019,  culliton_probing_2019} have explored correlations between intrinsic NAL velocity and \civ\ equivalent width distributions with quasar optical and radio luminosities at 4400 \AA\ and 5 GHz, respectively, as well as the radio loudness parameter, finding no clear trends.
We explore the correlations between associated NAL velocity, $v_{abs}$, and \civ\ equivalent width distributions with: the black hole mass $M_{\text{BH}, \mgii}$, bolometric luminosity, and the \civ\ blueshift. 
We find no clear trend with any of them, either considering the whole \civ\ or the restricted \civ+\nv\ NAL samples (Figure \ref{fig:NAL_dependence_on_quasar_properties}). 
Interestingly, in both samples, quasars with the highest bolometric luminosities show a lack of \nv\ absorbers associated with \civ. 
A possible, weak correlation may be present in the $EW_{\text{\civ,rf}}$ versus \civ\ blueshift plot, where one can see that as the \civ\ blueshift increases, there are progressively fewer \civ-only absorbers and a growing presence of \civ+\nv\ absorbers, whose \civ\ equivalent width concentrates around smaller values and with apparently less scatter.

\section{Summary and conclusions}
In this work, we analyse a sample of 39 luminous quasars ($ \log L_{\text{bol}}[\text{erg/s}] \gtrsim 47$) at redshift between 5 and 5.7, providing one of the few available studies -to our knowledge- in this critical yet mostly unexplored redshift range (see also \citealp{lai_xqz5_2024}).
Our analysis primarily leverages X-Shooter observations, supplemented with available ALMA archival data.
Using the X-Shooter spectra, we carry out a detailed investigation of absorption systems detected against the quasar rest-frame UV/optical emission. 
Our focus is on the high-ionisation (\civ+\nv) NAL systems within the quasar proximity zone, which are likely intrinsic and hence trace the quasar environment and outflows. 
To complement this, we investigate broad absorption lines (BALs), which are regarded as more robust outflow tracers than NALs. 
Furthermore, we compare the occurrence and properties of outflows derived from such absorption features with those retrieved from a complementary emission analysis, based on the examination of \mgii\ and \civ\ emission lines. 
In particular, the frequent asymmetric \civ\ line profiles and significant velocity shifts relative to \mgii\ strongly suggest the presence of powerful outflows. 
Therefore, our analysis provides clear evidence of ionised outflows, revealed through both absorption and emission measurements.
By combining these two complementary approaches, we aim to provide a comprehensive characterisation of high-redshift outflows and early feedback mechanisms occurring in the post-reionisation Universe. 
In order to further evaluate the effective impact of these quasar-driven winds on the host galaxies, we exploit the ALMA data to derive the galaxy dynamical masses through the study of the \cii$158 \mu$m emission line.

More quantitatively, our main findings can be summarised as follows: 
\begin{itemize}
    \item Associated \civ+\nv\ NAL systems and \civ\ BAL troughs are found, respectively, in 13 and 6 quasars\footnote{The fraction of BAL quasars in our sample cannot be considered as representative of the whole quasar population at $z \sim 5$, as the original sample was selected trying to avoid BALs. }, for a total of 18 sources with outflow detections in absorption. 
    Among these, J1004+2025 is the sole object of the sample displaying both an intrinsic NAL and a BAL.
    \item Black hole masses derived using the \mgii\ emission line span from $M_{\text{BH}, \text{\mgii}} \simeq 10^{8.7}$ to $10^{10.1} M_\odot$. 
    Bolometric luminosities vary in the range $\log (L_{\text{bol}} [\text{erg/s}]) \simeq 47-48$ , with median of $\sim 47.4$. 
    \item 35 out of 39 quasars exhibit \civ\ blueshifts exceeding 1000 km s$^{-1}$, with 22 surpassing 2000 km s$^{-1}$. 
    Regarding the \civ-\mgii\ velocity shift, 25 objects show $\Delta v$(\civ-\mgii) $< -1000$ km s$^{-1}$, and 14 of them even below -2000 km s$^{-1}$. 
    Such pronounced blueshifts and velocity shifts in the \civ\ line profile are commonly interpreted as strong outflow signatures. 
    This interpretation is further supported by the high Eddington ratios observed in the sample, with  32\% of the quasars lying in the super-Eddington regime.
    \item There is an apparent evolution of the average $\Delta v$(\civ-\mgii) with redshift (Fig. \ref{figure-evolution-of-vel-shift-with-redshift}), although it is still unclear whether this trend represents a real physical change in the quasar properties or just a selection bias affecting the highest-redshift quasar measurements.
    \item Our combined analysis of absorption and emission tracers is summarised in Fig. \ref{Figure-civ_fwhm-vs-vel-shift}, and reveals that both broad and narrow absorption features are present across the full range of \civ\ FWHM and $\Delta v$(\civ-\mgii) values. Notably, the fraction of quasars with ionised outflow detections in absorption increases with more extreme \civ-\mgii\ velocity shifts, exceeding $50$\% for $\Delta v$(\civ-\mgii) < -2000 km s$^{-1}$. 
    These results suggest that, while the physical mechanisms powering the two outflow phenomena detected in emission and absorption may differ, a correlation exists between them. It is plausible that they represent distinct phases and/or physical scales of the same outflow process. 
    This connection paves the way for new outflow studies centred on absorption lines, both broad and narrow. Such an approach is particularly valuable at high redshifts, where emission-based studies are often limited by data resolution and complexity. The potential of NALs as robust outflow tracers is particularly significant given their ubiquity in quasar spectra, in contrast to BALs that are found with an occurrence of $\sim 50$\% at $z \sim 6$ and even lower ($\sim 20$\%) at $ z \leq 4$ \citep{bischetti_fraction_2023}.
    \item Based on our limited sample, quasars at redshift $\sim 5$ exhibit a behaviour similar to those at $z \gtrsim 6$. 
    In particular, they still lie above the local $M_{\text{BH}}-M_{\text{host}}$ relation (Figure \ref{Figure-Mbh_Mdyn}), and display extremely high BAL velocities, up to $\sim 48.000$ km s$^{-1}$ (Table \ref{tab:BAL-properties}).  
    These findings suggest a phase of efficient black hole feedback occurring at redshift $\sim 6$ and likely persisting down to $z \sim 5$, characterised by strong outflows and by a rapid black hole growth exceeding that of the host galaxy. 
\end{itemize}

This study showcases the importance of investigating quasar properties in the $z \sim 5$ regime. 
The availability of more and higher-resolution data would allow a more detailed analysis of emission and absorption outflow properties, similar to that carried out in this work,  unraveling the early feedback mechanisms occurring in the post-reionisation Universe.

\begin{acknowledgements}
      This work is based on observations collected at the European Southern Observatory under ESO programs 098.A-01111 (PI: Rafelski, M) and 0100.A-0243 (PI: Rafelski, M), and makes use of the following ALMA data: 2019.1.00840.S, 2019.1.01721.S, 2022.1.00662.S.
      The research activities described in this paper were carried out with contribution of the Next Generation EU funds within the National Recovery and Resilience Plan (PNRR), Mission 4 - Education and Research, Component 2 - From Research to Business (M4C2), Investment Line 3.1 - Strengthening and creation of Research Infrastructures, Project IR0000034 – ``STILES - Strengthening the Italian Leadership in ELT and SKA''. 
      C.F. acknowledges financial support from PRIN MUR 2022 grant: project 2022TKPB2P - BIG-z; from Ricerca Fondamentale INAF 2023 Data Analysis grant ``ARCHIE ARchive Cosmic HI \& ISM  Evolution''; and from the M4C2 Missione 4 “Istruzione e Ricerca” - Componente C2 Investimento 1.1 Fondo per il Programma Nazionale di Ricerca e Progetti di Rilevante Interesse Nazionale (PRIN) prog. PRIN 2022 PNRR P2022ZLW4T, "Next-generation computing and data technologies to probe the cosmic metal content". 
      M.B. acknowledges support from INAF under project 1.05.12.04.01 - 431 MINI-GRANTS di RSN1 "Mini-feedback"  and support from UniTs under project DF-microgrants23 "Hyper-Gal". 
      V.D. acknowledges financial support from the Bando Ricerca Fondamentale INAF 2022 Large Grant “XQR-30”.
\end{acknowledgements}

\bibliographystyle{aa}
\bibliography{aa54010-25}

\clearpage
\appendix

\section{Properties of absorption systems}

\begin{minipage}{\textwidth}
    \centering
    \captionof{table}{Properties of our sample of \civ\ absorption systems, including quasar name and emission redshift, redshift and velocity separation of the absorber, its rest-frame \civ\ equivalent width, and \civ\ column density. For each \civ\ absorber, we list the column densities of associated \nv\ and/or \siiv\ absorptions, if detected. The emission redshift is estimated from \mgii\ or \cii, if available. The complete table is available online as supplementary material.} \label{table_absorption_systs_properties}
    \begin{tabular}{cccccccc}
\toprule
 Quasar & $z_{em}$ & $z_{abs}$ & $v_{abs}$ & $EW_{\text{\civ,rf}}$ & $\log (N_{\text{\civ}})$ & $\log (N_{\text{\nv}})$ & $\log (N_{\text{\siiv}})$\\[5pt]
 & & & [km s$^{-1}$] & [\AA] & [cm$^{-2}$] & [cm$^{-2}$] & [cm$^{-2}$]\\[5pt]
\midrule
J0251+0333 & 4.9900 & $4.804$ & $9446$ & $0.06 \pm 0.04$ & $13.71 \pm 0.15$ & \textemdash  & \textemdash\\[5pt]
J0251+0333 & 4.9900 & $4.865$ & $6343$ & $0.55 \pm 0.02$ & $13.65 \pm 0.15$ & \textemdash  & \textemdash\\[5pt]
J0251+0333 & 4.9900 & $4.869$ & $6128$ & $0.63 \pm 0.02$ & $14.47 \pm 0.14$ & \textemdash  & \textemdash\\[5pt]
J0251+0333 & 4.9900 & $4.899$ & $4601$ & $0.40 \pm 0.02$ & $13.78 \pm 0.17$ & $13.76 \pm 0.14$  & \textemdash\\[5pt]
J0251+0333 & 4.9900 & $4.936$ & $2727$ & $0.15 \pm 0.02$ & $13.93 \pm 0.17$ & \textemdash  & \textemdash\\[5pt]
J1200+1817 & 5.0000 & $4.858$ & $7162$ & $0.05 \pm 0.02$ & $13.15 \pm 0.14$ & \textemdash  & \textemdash\\[5pt]
J0221-0342 & 5.0170 & $4.830$ & $9460$ & $0.01 \pm 0.05$ & $13.27 \pm 0.09$ & \textemdash  & \textemdash\\[5pt]
J0017-1000 & 5.0170 & $4.823$ & $9822$ & $0.32 \pm 0.02$ & $13.93 \pm 0.08$ & \textemdash  & \textemdash\\[5pt]
J0338+0021 & 5.0230 & $4.977$ & $2278$ & $0.08 \pm 0.03$ & $13.26 \pm 0.09$ & \textemdash  & \textemdash\\[5pt]
J1423+1303 & 5.0462 & $4.868$ & $8969$ & $0.12 \pm 0.02$ & $13.41 \pm 0.08$ & \textemdash  & \textemdash\\[5pt]
J1423+1303 & 5.0462 & $5.017$ & $1440$ & $1.39 \pm 0.03$ & $14.19 \pm 0.08$ & $14.27 \pm 0.07$  & \textemdash\\[5pt]
  & & & & ... & & & \\[5pt]
\bottomrule
    \end{tabular}
\end{minipage}

\clearpage
\newpage

\section{Quasar properties from UV emission lines}

\begin{minipage}{\textwidth}
    \centering
    \captionof{table}{Emission properties of \mgii\ and \civ, along with the derived quasar physical quantities: redshift, lines FWHM, \civ\ blueshift, \civ--\mgii\ velocity shift, monochromatic luminosity at 3000\AA, bolometric luminosity, black hole mass, and Eddington ratio.}
    \label{tab:quasar_properties}
    \adjustbox{max width=1.0\textwidth}{
    \begin{tabular}{cccccccccccc}
    \toprule
 Quasar & $z_\mgii$ & FWHM$_{\text{\civ}}$ & FWHM$_{\text{\mgii}}$ & \civ\ blueshift & $\Delta v$(\civ-\mgii) & $\log \lambda L_{3000 \text{\AA}}$ & $\log L_{\text{bol}}$ & $\log M_{\text{BH, \civ}}$ & $\log M_{\text{BH, \mgii}}$  & $\lambda_{\text{Edd, \civ}}$ & $\lambda_{\text{Edd, \mgii}}$\\[5pt]
 & & [km/s] & [km/s] & [km/s] & [km/s] & [erg s$^{-1}$] & [erg s$^{-1}$] & [$M_\odot$] & [$M_\odot$] & & \\[5pt]\midrule
J0251+0333$^{*}$ & $4.9900 \pm 0.0008$ & $4430 \pm 130$ & $2590 \pm 60$ & $1250 \pm 60$ & $-410 \pm 40$ & $46.79 \pm 0.05$ & $47.50 \pm 0.05$ &  $9.35 \pm 0.04$ & $9.08 \pm 0.03$ & $1.1\pm 0.2$ & $2.0\pm 0.3$\\[5pt]
J2344+1653$^{*}$ & $4.9980 \pm 0.0003$ & $6850 \pm 860$ & $3530 \pm 240$ & $-230 \pm 140$ & $-350 \pm 200$ & $47.04 \pm 0.12$ & $47.75 \pm 0.12$ &  $10.47 \pm 0.14$ & $9.47 \pm 0.08$ & $0.1\pm 0.1$ & $1.4\pm 0.5$\\[5pt]
J1200+1817$^{*}$ & $5.0000 \pm 0.0010$ & $5970 \pm 90$ & $2960 \pm 110$ & $1720 \pm 40$ & $-1460 \pm 50$ & $46.60 \pm 0.13$ & $47.32 \pm 0.13$ &  $9.43 \pm 0.07$ & $9.10 \pm 0.07$ & $0.6\pm 0.2$ & $1.2\pm 0.4$\\[5pt]
J0221-0342$^{*}$ & $5.0170 \pm 0.0008$ & $2750 \pm 230$ & $2200 \pm 80$ & $1050 \pm 60$ & $-370 \pm 80$ & $46.73 \pm 0.11$ & $47.44 \pm 0.11$ &  $8.94 \pm 0.36$ & $8.91 \pm 0.06$ & $2.4\pm 2.1$ & $2.6\pm 0.8$\\[5pt]
J0017-1000$^{*}$ & $5.0170 \pm 0.0020$ & $5810 \pm 100$ & $3420 \pm 210$ & $2330 \pm 60$ & $-2070 \pm 100$ & $46.63 \pm 0.10$ & $47.35 \pm 0.10$ &  $9.24 \pm 0.03$ & $9.24 \pm 0.07$ & $0.6\pm 1.1$ & $1.0\pm 0.3$\\[5pt]
J0846+0800$^{*}$ & $5.0210 \pm 0.0016$ & $8680 \pm 190$ & $3110 \pm 130$ & $3400 \pm 100$ & $-3140 \pm 90$ & $46.52 \pm 0.14$ & $47.23 \pm 0.14$ &  $9.37 \pm 0.82$ & $9.10 \pm 0.08$ & $1.0\pm 0.2$ & $1.0\pm 0.4$\\[5pt]
J0338+0021$^{*}$ & $5.0230 \pm 0.0005$ & $7200 \pm 1830$ & $3510 \pm 270$ & $3280 \pm 1010$ & $-1560 \pm 1720$ & $46.44 \pm 0.22$ & $47.15 \pm 0.22$ &  $9.22 \pm 0.30$ & $9.17 \pm 0.13$ & $0.7\pm 0.6$ & $0.7\pm 0.4$\\[5pt]
J1423+1303$^{*}$ & $5.0462 \pm 0.0018$ & $7590 \pm 100$ & $4030 \pm 270$ & $2330 \pm 40$ & $-2070 \pm 80$ & $46.68 \pm 0.15$ & $47.39 \pm 0.15$ &  $9.54 \pm 0.02$ & $9.41 \pm 0.09$ & $10.1\pm 6.0$ & $0.7\pm 0.3$\\[5pt]
J1601-1828$^{*}$ & $5.0502 \pm 0.0006$ & $6180 \pm 580$ & $3560 \pm 140$ & $4210 \pm 310$ & $-2760 \pm 270$ & $46.70 \pm 0.04$ & $47.41 \pm 0.04$ &  $8.96 \pm 0.11$ & $9.31 \pm 0.04$ & $0.5\pm 0.2$ & $1.0\pm 0.1$\\[5pt]
J0025-0145$^{*}$ & $5.0630 \pm 0.0030$ & $5500 \pm 1480$ & $2310 \pm 40$ & $8080 \pm 920$ & $-920 \pm 0$ & $47.19 \pm 0.07$ & $47.90 \pm 0.07$ &  $8.78 \pm 0.25$ & $9.18 \pm 0.04$ & $0.8\pm 0.6$ & $4.0\pm 0.7$\\[5pt]
J1004+2025$^{*}$ & $5.0634 \pm 0.0014$ & $14830 \pm 1880$ & $5450 \pm 350$ & $5030 \pm 250$ & $-3160 \pm 320$ & $46.53 \pm 0.24$ & $47.24 \pm 0.24$ &  $9.57 \pm 0.27$ & $9.60 \pm 0.13$ & $2.2\pm 0.6$ & $0.3\pm 0.2$\\[5pt]
J0835+0537 & $5.0643 \pm 0.0010$ & $7130 \pm 150$ & $3590 \pm 130$ & $4210 \pm 110$ & $-4000 \pm 120$ & $46.47 \pm 0.14$ & $47.18 \pm 0.14$ &  $9.02 \pm 0.03$ & $9.20 \pm 0.08$ & $1.1\pm 0.4$ & $0.7\pm 0.3$\\[5pt]
J2202+1509 & $5.0755 \pm 0.0007$ & $5550 \pm 290$ & $4370 \pm 620$ & $1630 \pm 80$ & $-1380 \pm 220$ & $46.91 \pm 0.33$ & $47.62 \pm 0.33$ &  $9.60 \pm 0.05$ & $9.60 \pm 0.20$ & $0.4\pm 0.3$ & $0.8\pm 0.7$\\[5pt]
J0115-0253$^{*}$ & $5.0820 \pm 0.0020$ & $5900 \pm 320$ & $4720 \pm 80$ & $2210 \pm 90$ & $-2050 \pm 100$ & $46.49 \pm 0.19$ & $47.20 \pm 0.19$ &  $9.27 \pm 0.08$ & $9.45 \pm 0.10$ & $0.7\pm 0.3$ & $0.4\pm 0.2$\\[5pt]
J2201+0302$^{*}$ & $5.0887 \pm 0.0025$ & $6360 \pm 200$ & $4250 \pm 200$ & $3280 \pm 70$ & $-2280 \pm 90$ & $46.71 \pm 0.19$ & $47.42 \pm 0.19$ &  $9.25 \pm 0.03$ & $9.47 \pm 0.10$ & $1.8\pm 0.9$ & $0.7\pm 0.3$\\[5pt]
J2226-0618$^{*}$ & $5.1050 \pm 0.0010$ & $5630 \pm 1090$ & $3460 \pm 220$ & $2380 \pm 270$ & $-1600 \pm 310$ & $47.07 \pm 0.10$ & $47.78 \pm 0.10$ &  $9.40 \pm 0.18$ & $9.47 \pm 0.07$ & $1.1\pm 0.5$ & $1.6\pm 0.4$\\[5pt]
J0957+1016$^{*}$ & $5.1274 \pm 0.0011$ & $7750 \pm 180$ & $4890 \pm 750$ & $2620 \pm 90$ & $-2350 \pm 240$ & $46.37 \pm 0.14$ & $47.08 \pm 0.14$ &  $9.37 \pm 0.09$ & $9.42 \pm 0.15$ & $0.4\pm 0.2$ & $0.3\pm 0.2$\\[5pt]
J0957+0610$^{*}$ & $5.1650 \pm 0.0030$ & $7210 \pm 200$ & $4190 \pm 430$ & $2090 \pm 80$ & $-1890 \pm 180$ & $46.76 \pm 0.14$ & $47.47 \pm 0.14$ &  $9.58 \pm 0.03$ & $9.48 \pm 0.12$ & $0.6\pm 0.2$ & $0.7\pm 0.3$\\[5pt]
J0854+2056$^{*}$ & $5.1715 \pm 0.0007$ & $2210 \pm 100$ & $2450 \pm 540$ & $790 \pm 20$ & $20 \pm 70$ & $46.69 \pm 0.12$ & $47.40 \pm 0.12$ &  $8.84 \pm 0.04$ & $8.98 \pm 0.20$ & $2.8\pm 0.9$ & $2.0\pm 1.1$\\[5pt]
J0241+0435$^{*}$ & $5.1900 \pm 0.0006$ & $4080 \pm 380$ & $3260 \pm 140$ & $1160 \pm 70$ & $-540 \pm 80$ & $46.65 \pm 0.12$ & $47.36 \pm 0.12$ &  $9.25 \pm 1.48$ & $9.21 \pm 0.07$ & $1.0\pm 0.2$ & $1.1\pm 0.4$\\[5pt]
J0131-0321$^{*}$ & $5.1942 \pm 0.0007$ & $6400 \pm 450$ & $2750 \pm 70$ & $1400 \pm 160$ & $-830 \pm 160$ & $47.32 \pm 0.07$ & $48.04 \pm 0.07$ &  $9.92 \pm 0.08$ & $9.40 \pm 0.04$ & $1.0\pm 3.4$ & $3.3\pm 0.6$\\[5pt]
J0902+0851$^{*}$ & $5.2200 \pm 0.0008$ & $2520 \pm 110$ & $2070 \pm 220$ & $470 \pm 30$ & $-20 \pm 50$ & $46.40 \pm 0.19$ & $47.11 \pm 0.19$ &  $8.97 \pm 0.05$ & $8.69 \pm 0.13$ & $1.1\pm 0.5$ & $2.0\pm 1.1$\\[5pt]
J0216+2304$^{*}$ & $5.2273 \pm 0.0005$ & $5960 \pm 690$ & $3930 \pm 190$ & $2700 \pm 220$ & $-1720 \pm 240$ & $46.67 \pm 0.12$ & $47.39 \pm 0.12$ &  $9.20 \pm 0.13$ & $9.39 \pm 0.07$ & $0.8\pm 0.7$ & $0.8\pm 0.3$\\[5pt]
J2325-0553$^{*}$ & $5.2304 \pm 0.0005$ & $7170 \pm 2290$ & $4870 \pm 400$ & $2700 \pm 560$ & $-1570 \pm 1800$ & $46.76 \pm 0.19$ & $47.47 \pm 0.19$ &  $9.45 \pm 0.30$ & $9.61 \pm 0.12$ & $1.2\pm 0.5$ & $0.6\pm 0.3$\\[5pt]
J1436+2132$^{*}$ & $5.2356 \pm 0.0005$ & $5500 \pm 970$ & $3420 \pm 110$ & $1830 \pm 210$ & $-870 \pm 600$ & $46.45 \pm 0.26$ & $47.16 \pm 0.26$ &  $9.30 \pm 0.17$ & $9.15 \pm 0.13$ & $1.0\pm 0.1$ & $0.8\pm 0.5$\\[5pt]
J2351-0459$^{*}$ & $5.2481 \pm 0.0009$ & $5180 \pm 850$ & $3780 \pm 120$ & $1400 \pm 270$ & $-680 \pm 270$ & $46.46 \pm 0.17$ & $47.17 \pm 0.17$ &  $9.34 \pm 0.27$ & $9.24 \pm 0.09$ & $0.5\pm 0.4$ & $0.7\pm 0.3$\\[5pt]
J2225+0330$^{*}$ & $5.2523 \pm 0.0004$ & $4780 \pm 560$ & $5450 \pm 100$ & $1600 \pm 250$ & $-740 \pm 110$ & $46.49 \pm 0.18$ & $47.20 \pm 0.18$ &  $9.17 \pm 0.12$ & $9.58 \pm 0.09$ & $0.6\pm 0.4$ & $0.3\pm 0.1$\\[5pt]
J1147-0109$^{*}$ & $5.2544 \pm 0.0006$ & $9070 \pm 210$ & $5680 \pm 260$ & $1150 \pm 90$ & $-890 \pm 100$ & $46.77 \pm 0.28$ & $47.48 \pm 0.28$ &  $10.04 \pm 0.04$ & $9.75 \pm 0.15$ & $0.8\pm 0.4$ & $0.4\pm 0.3$\\[5pt]
J0747+1153 & $5.2575 \pm 0.0004$ & $5050 \pm 140$ & $3040 \pm 60$ & $1050 \pm 20$ & $-840 \pm 30$ & $47.19 \pm 0.03$ & $47.90 \pm 0.03$ &  $9.77 \pm 0.03$ & $9.42 \pm 0.02$ & $0.2\pm 0.1$ & $2.3\pm 0.2$\\[5pt]
J2330+0957 & $5.2889 \pm 0.0007$ & $11390 \pm 2840$ & $4260 \pm 420$ & $3140 \pm 680$ & $-3010 \pm 630$ & $46.61 \pm 0.12$ & $47.32 \pm 0.12$ &  $9.64 \pm 1.04$ & $9.43 \pm 0.10$ & $1.0\pm 0.6$ & $0.6\pm 0.2$\\[5pt]
J2358+0634 & $5.2973 \pm 0.0007$ & $7200 \pm 960$ & $3360 \pm 380$ & $3570 \pm 250$ & $-2660 \pm 350$ & $46.70 \pm 0.23$ & $47.41 \pm 0.23$ &  $9.28 \pm 0.12$ & $9.26 \pm 0.15$ & $0.4\pm 0.9$ & $1.1\pm 0.7$\\[5pt]
J0812+0440 & $5.3026 \pm 0.0009$ & $3000 \pm 70$ & $2400 \pm 60$ & $670 \pm 20$ & $-180 \pm 30$ & $46.58 \pm 0.09$ & $47.30 \pm 0.09$ &  $9.12 \pm 0.03$ & $8.91 \pm 0.05$ & $1.1\pm 0.2$ & $1.9\pm 0.4$\\[5pt]
J0116+0538 & $5.3310 \pm 0.0004$ & $6290 \pm 540$ & $3910 \pm 190$ & $2330 \pm 120$ & $-1880 \pm 250$ & $46.74 \pm 0.24$ & $47.45 \pm 0.24$ &  $9.42 \pm 0.08$ & $9.41 \pm 0.13$ & $0.8\pm 0.5$ & $0.8\pm 0.5$\\[5pt]
J0155+0415 & $5.3596 \pm 0.0011$ & $8890 \pm 1730$ & $4340 \pm 390$ & $4590 \pm 540$ & $-3480 \pm 770$ & $46.70 \pm 0.21$ & $47.41 \pm 0.21$ &  $9.28 \pm 0.19$ & $9.48 \pm 0.13$ & $1.0\pm 0.7$ & $0.6\pm 0.4$\\[5pt]
J0306+1853 & $5.3649 \pm 0.0005$ & $7850 \pm 350$ & $5690 \pm 1330$ & $2790 \pm 130$ & $-2460 \pm 1490$ & $47.52 \pm 0.25$ & $48.23 \pm 0.25$ &  $9.92 \pm 0.05$ & $10.13 \pm 0.24$ & $1.6\pm 0.9$ & $1.0\pm 0.8$\\[5pt]
J1022+2252$^{(a)}$ & $5.47 \pm 0.01^{(b)}$ & $3410 \pm 340$ & \textemdash & $1130 \pm 100$ & \textemdash & $46.80 \pm 0.09$ & $47.51 \pm 0.09$ &  $9.18 \pm 0.09$ & \textemdash & $1.7\pm 0.5$ & \textemdash\\[5pt]
J2207-0416$^{(a)}$ & $5.523 \pm 0.004$ & $6300 \pm 1210$ & $4070 \pm 240$ & $1250 \pm 330$ & $-990 \pm 270$ & $47.10 \pm 0.06$ & $47.81 \pm 0.06$ &  $9.74 \pm 0.19$ & $9.63 \pm 0.06$ & $0.9\pm 0.4$ & $1.2\pm 0.2$\\[5pt]
J0108+0711$^{(a)}$ & $5.5578 \pm 0.0018$ & $5040 \pm 460$ & $2770 \pm 1120$ & $2590 \pm 170$ & $-2320 \pm 260$ & $46.74 \pm 0.14$ & $47.45 \pm 0.14$ &  $9.18 \pm 0.09$ & $9.11 \pm 0.36$ & $1.5\pm 0.6$ & $1.7\pm 1.5$\\[5pt]
J1335-0328$^{(a)}$ & $5.699 \pm 0.001^{(b)}$ & $4650 \pm 120$ & \textemdash & $2090 \pm 60$ & \textemdash & $46.83 \pm 0.21$ & $47.55 \pm 0.21$ &  $9.34 \pm 0.03$ & \textemdash & $1.2\pm 0.6$ & \textemdash\\[5pt]
\bottomrule
\end{tabular}}
\parbox{\textwidth}{
\vspace{2mm}
\footnotesize
\textbf{Notes:} 
    Quasars marked with an asterisk are in common with the XQz5 sample described in \cite{lai_xqz5_2024}.
    The reported errors account only for the statistical uncertainties. Black hole masses and bolometric luminosities are also affected by systematic uncertainties of 0.5 dex and 0.3 dex, respectively.
    
    \textit{(a)} \mgii\ emission line contaminated by telluric emission. 
    \textit{(b)} Redshift estimated from the apparent start of the Ly$\alpha$ forest \citep{becker_evolution_2019}.
    \textit{(c)} Redshift estimated from the  \cii\ emission line (Table \ref{table-ALMA-cii-lines}). }

\end{minipage}

\clearpage
\newpage 

\section{Comparison of emission line profiles}

\begin{center}
\begin{minipage}{\textwidth}
    \centering
    \includegraphics[width = 0.8\textwidth]{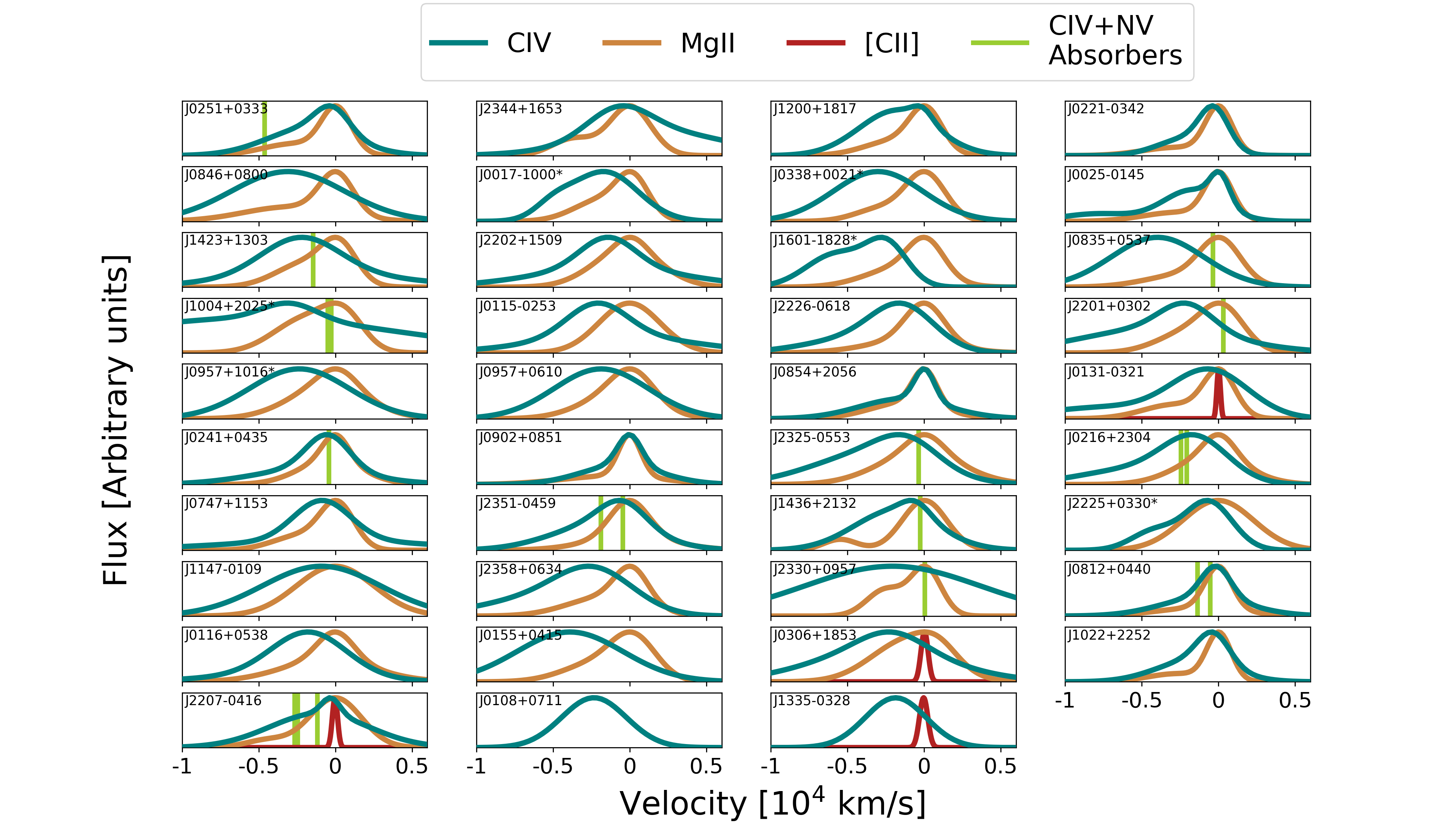}
    \captionof{figure}{Comparison between \civ, \mgii\ and \cii\ emission line profiles for each quasar of our sample. The green vertical lines highlight the presence of \civ+\nv\ absorbers.} 
    \label{fig:comparison_line_profiles}
\end{minipage}
\end{center}


\section{Connecting intrinsic NAL and quasar properties}
\begin{center}
\begin{minipage}{\textwidth}
    \centering
    \includegraphics[width =0.7\textwidth]{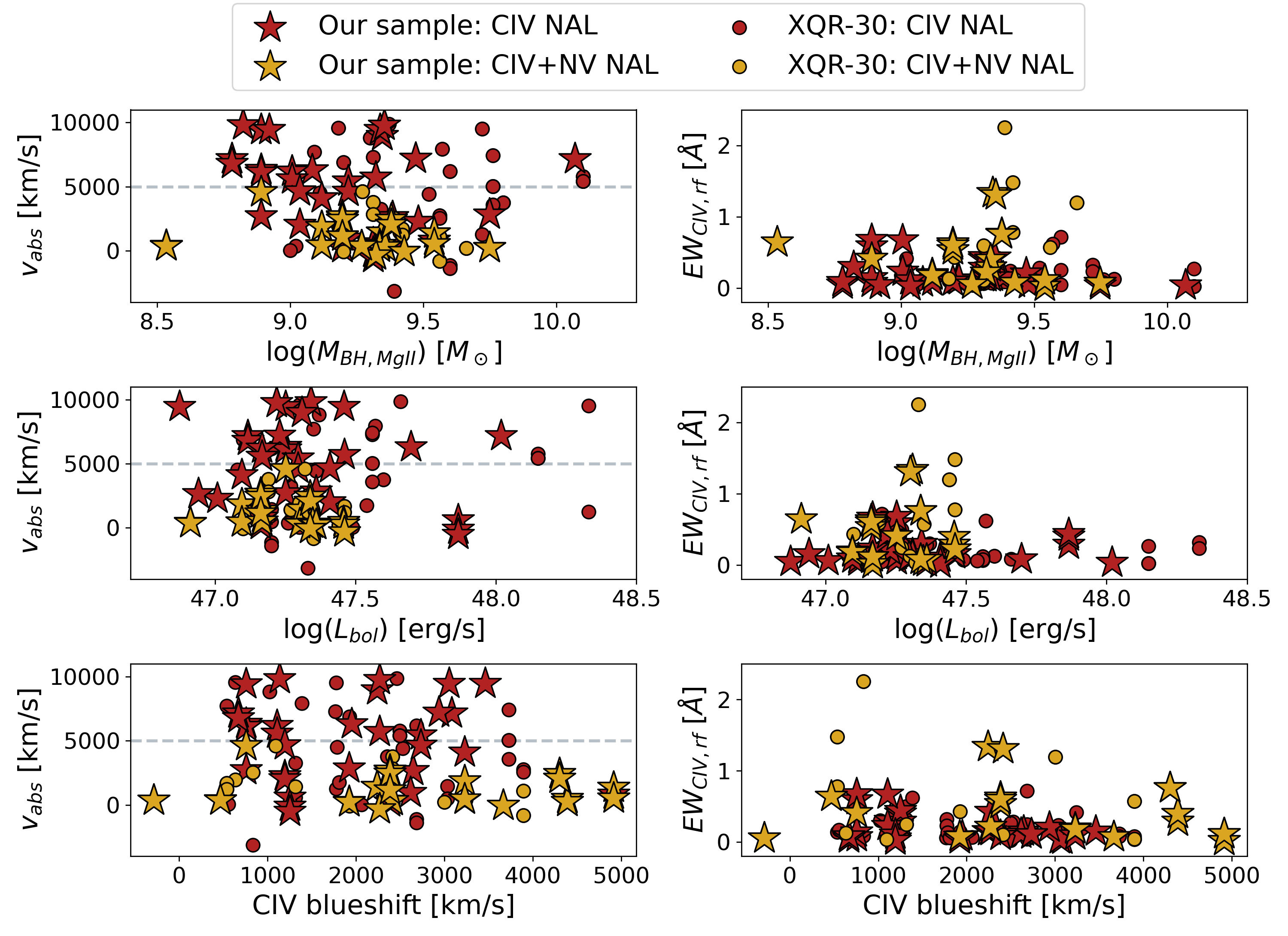}
     \captionof{figure}{Absorption system velocities, $v_{abs}$, and \civ\ equivalent widths versus quasar properties.
     From top to bottom, black hole mass, bolometric luminosity, and \civ\ blueshift.
    Golden and red stars (dots) correspond, respectively, to \civ+\nv\ and \civ-only absorption systems from our (XQR-30) sample. 
    Narrow absorbers and black hole properties do not show evident correlations. An interesting result is the lack of \nv\ absorption systems at the highest bolometric luminosities. }
    \label{fig:NAL_dependence_on_quasar_properties}
\end{minipage}
\end{center}
\clearpage

\end{document}